\documentclass[review,3p,times,authoryear]{elsarticle}

\usepackage{amsmath,amssymb}
\usepackage{graphics,graphicx,epsfig}
\usepackage{natbib}
\usepackage{tabularx,array,multirow,multicol,booktabs}
\usepackage{color}
\usepackage{siunitx}
\usepackage{orcidlink}
\usepackage{hyperref}
\usepackage{lineno}
\numberwithin{equation}{section}



\journal{arXiv: Soft Condensed Matter}

\begin{document}
%
%
\newcommand{\calA}{\mathcal{A}}%
\newcommand{\calB}{\mathcal{B}}%
\newcommand{\calC}{\mathcal{C}}%
\newcommand{\calD}{\mathcal{D}}%
\newcommand{\calE}{\mathcal{E}}%
\newcommand{\calF}{\mathcal{F}}%
\newcommand{\calH}{\mathcal{H}}%
\newcommand{\calI}{\mathcal{I}}%
\newcommand{\calJ}{\mathcal{J}}%
\newcommand{\calK}{\mathcal{K}}%
\newcommand{\calL}{\mathcal{L}}%
\newcommand{\calM}{\mathcal{M}}%
\newcommand{\calN}{\mathcal{N}}%
\newcommand{\calP}{\mathcal{P}}%
\newcommand{\calQ}{\mathcal{Q}}%
\newcommand{\calR}{\mathcal{R}}%
\newcommand{\calS}{\mathcal{S}}%
\newcommand{\calV}{\mathcal{V}}%
\newcommand{\calW}{\mathcal{W}}%
\newcommand{\calX}{\mathcal{X}}%
\newcommand{\calT}{\mathcal{T}}%
\newcommand{\bfa}{{\bf a}}\newcommand{\bfA}{{\bf A}}%
\newcommand{\bfb}{{\bf b}}\newcommand{\bfB}{{\bf B}}%
\newcommand{\bfc}{{\bf c}}\newcommand{\bfC}{{\bf C}}%
\newcommand{\bfd}{{\bf d}}\newcommand{\bfD}{{\bf D}}%
\newcommand{\bfe}{{\bf e}}\newcommand{\bfE}{{\bf E}}%
\newcommand{\bff}{{\bf f}}\newcommand{\bfF}{{\bf F}}%
\newcommand{\bfg}{{\bf g}}\newcommand{\bfG}{{\bf G}}%
\newcommand{\bfh}{{\bf h}}\newcommand{\bfH}{{\bf H}}%
\newcommand{\bfi}{{\bf i}}\newcommand{\bfI}{{\bf I}}%
\newcommand{\bfj}{{\bf j}}\newcommand{\bfJ}{{\bf J}}%
\newcommand{\bfk}{{\bf k}}\newcommand{\bfK}{{\bf K}}%
\newcommand{\bfl}{{\bf l}}\newcommand{\bfL}{{\bf L}}%
\newcommand{\bfm}{{\bf m}}\newcommand{\bfM}{{\bf M}}%
\newcommand{\bfn}{{\bf n}}\newcommand{\bfN}{{\bf N}}%
\newcommand{\bfo}{{\bf o}}\newcommand{\bfO}{{\bf O}}%
\newcommand{\bfp}{{\bf p}}\newcommand{\bfP}{{\bf P}}%
\newcommand{\bfq}{{\bf q}}\newcommand{\bfQ}{{\bf Q}}%
\newcommand{\bfr}{{\bf r}}\newcommand{\bfR}{{\bf R}}%
\newcommand{\bfs}{{\bf s}}\newcommand{\bfS}{{\bf S}}%
\newcommand{\bft}{{\bf t}}\newcommand{\bfT}{{\bf T}}%
\newcommand{\bfu}{{\bf u}}\newcommand{\bfU}{{\bf U}}%
\newcommand{\bfv}{{\bf v}}\newcommand{\bfV}{{\bf V}}%
\newcommand{\bfw}{{\bf w}}\newcommand{\bfW}{{\bf W}}%
\newcommand{\bfx}{{\bf x}}\newcommand{\bfX}{{\bf X}}%
\newcommand{\bfy}{{\bf y}}\newcommand{\bfY}{{\bf Y}}%
\newcommand{\bfz}{{\bf z}}\newcommand{\bfZ}{{\bf Z}}%
\newcommand{\bfalpha}{\bm{\alpha}}%
\newcommand{\bfbeta}{\bm{\beta}}%
\newcommand{\bfgamma}{\bm{\gamma}}%
\newcommand{\bfGamma}{\boldsymbol{\Gamma}}%
\newcommand{\bfdelta}{\bm{\delta}}%
\newcommand{\bfepsilon}{\bm{\epsilon}}%
\newcommand{\bfzeta}{\bm{\zeta}}%
\newcommand{\bfeta}{\bm{\eta}}%
\newcommand{\bftheta}{\bm{\theta}}%
\newcommand{\bfiota}{\bm{\iota}}%
\newcommand{\bfkappa}{\bm{\kappa}}%
\newcommand{\bflambda}{\bm{\lambda}}%
\newcommand{\bfLambda}{\boldsymbol{\Lambda}}%
\newcommand{\bfmu}{\bm{\mu}}%
\newcommand{\bfnu}{\bm{\nu}}%
\newcommand{\bfxia}{\bm{\xi}}%
\newcommand{\bfpi}{\bm{\pi}}%
\newcommand{\bfrho}{\bm{\rho}}%
\newcommand{\bfsigma}{\bm{\sigma}}%
\newcommand{\bftau}{\bm{\tau}}%
\newcommand{\bfupsilon}{\bm{\upsilon}}%
\newcommand{\bfphi}{\bm{\phi}}%
\newcommand{\bfPhi}{\bm{\Phi}}%
\newcommand{\bfchi}{\boldsymbol{\chi}}%
\newcommand{\bfpsi}{\bm{\psi}}%
\newcommand{\bfomega}{\bm{\omega}}%
\newcommand{\bfvarepsilon}{\bm{\varepsilon}}%
\newcommand{\bfvartheta}{\bm{\vartheta}}%
\newcommand{\bfvarpi}{\bm{\vaarphi}}%
\newcommand{\bfvarrho}{\bm{\varrho}}%
\newcommand{\bfvarsigma}{\bm{\varsigma}}%
\newcommand{\bfvarkappa}{\bm{\varkappa}}%
\newcommand{\bfvarphi}{\bm{\varphi}}%
\newcommand{\Det}{\hbox{\rm det}\mskip2mu}
\newcommand{\cof}{\hbox{\rm cof}\mskip2mu}
\newcommand{\skw}{\hbox{\rm skw}\mskip2mu}
\newcommand{\sym}{\hbox{\rm sym}\mskip2mu}
\newcommand{\symz}{\text{sym}_0\mskip2mu}
\newcommand{\Tr}{\hbox{\rm tr}\mskip2mu}
\newcommand{\dev}{\text{dev}\mskip2mu}
\newcommand{\Rot}{\hbox{\rm Rot}\mskip2mu}
\newcommand{\Lin}{\hbox{\rm Lin}}
\newcommand{\Orth}{\hbox{\rm Orth}}
\newcommand{\rot}{\text{Orth}^+}
\newcommand{\trans}{\scriptscriptstyle\top}
\newcommand{\zed}{{\bf 0}}
\newcommand{\id}{{\bf 1}}
\newcommand{\Grad}{\hbox{\rm grad}\mskip2mu}
\newcommand{\Div}{\hbox{\rm Div}\mskip2mu}
\newcommand{\curl}{\hbox{\rm curl}\mskip2mu}
\newcommand{\Curl}{\hbox{\rm Curl}\mskip2mu}
\newcommand{\divx}{\text{div}\mskip2mu}
\newcommand{\grad}{\text{grad}\mskip2mu}
\newcommand{\B}{\text{B}}
\newcommand{\Bt}{\calB_t}
\newcommand{\dBt}{\partial\calB_t}
\newcommand{\p}{\text{P}}
\newcommand{\pt}{\calP_t}
\newcommand{\dpt}{\partial\calP_t}

\begin{frontmatter}
\title{Large-deformation constitutive modeling of viscoelastic foams: Application to an open-cell foam material}
\author[1]{Jialiang Tao\orcidlink{0000-0003-4846-0631}\corref{cor2}}
\author[1,2,3,4]{Alexander K. Landauer\orcidlink{0000-0003-2863-039X}\corref{cor2}}
\author[2]{Zicheng Yan\orcidlink{0009-0003-2054-4697}\corref{cor2}}
\author[2]{Xiuqi Li}
\author[1]{Christian Franck\orcidlink{0000-0002-2347-620X}}
\author[2]{David L. Henann\orcidlink{0000-0002-1497-4143}\corref{cor1}}
\cortext[cor1]{Corresponding authors. E-mail addresses: alexander\_landauer@alumni.brown.edu (A.K. Landauer), david\_henann@brown.edu (D.L. Henann)}
\cortext[cor2]{These authors contributed equally to this work.}
\address[1]{Department of Mechanical Engineering, University of Wisconsin-Madison, Madison, WI, USA}
\address[2]{School of Engineering, Brown University, Providence, RI, USA}
\address[3]{Material Measurement Laboratory, National Institute of Standards and Technology, Gaithersburg, MD, USA}
\address[4]{Department of Chemistry \& Biochemistry, University of Maryland College Park, College Park, MD, USA}

\begin{abstract}
Open-cell elastomeric foams are widely utilized for impact protection and cushioning applications. These foams can demonstrate remarkable elastic compressibility and rate sensitivity over a wide range of force and acceleration protection levels. They often exhibit volumetric--distortional coupling in their mechanical response and potentially compaction banding when subjected to dynamic loads. Here, a phenomenological modeling framework, adopted previously for closed-cell foams, is specified using an experimental characterization-based approach for calibrating and validating the large-deformation, viscoelastic material model. Homogeneous simple compression and tension experiments were performed on isotropic, open-cell viscoelastic foam materials, and stress--strain and lateral--axial strain data were used for model calibration. Data were collected at decadal strain rates to inform the rate-sensitive aspects of the model. The modeling approach involved decomposing the total material response into an equilibrium hyperelastic contribution and several nonlinear, Maxwell-like contributions and utilized invariants of the logarithmic strain to account for large deformations and the strong volumetric--distortional coupling effect. A new strain localization detection algorithm was developed and used to check for the formation of compaction bands, which would invalidate modeling assumptions for the calibration experiments. To validate the predictive capability of the constitutive model, experiments were conducted that involved intentionally inhomogeneous deformations over a spectrum of strain rates and deformation types. The consistency between the measured responses from the validation experiments and the model predictions obtained from finite-element-based numerical simulations was largely within the experimental uncertainty bounds, demonstrating the predictive capability of the model. 

\end{abstract}


\end{frontmatter}

\section{Introduction}\label{sec:intro}
\begin{figure}[!t]
	\centerline{\includegraphics[width=0.95\textwidth]{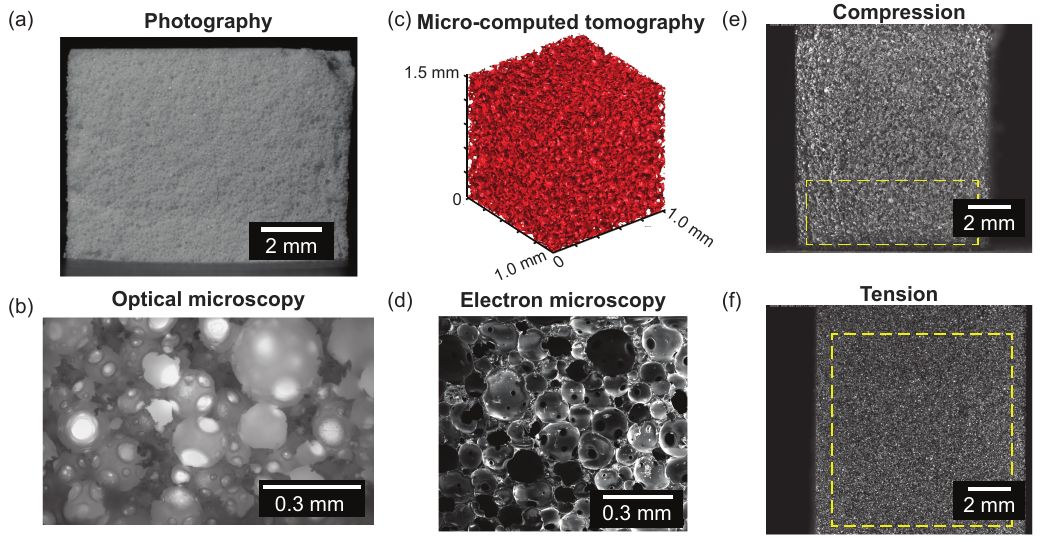}}
	\caption{Images of an open-cell viscoelastic foam at various imaging scales. (a) Photographic, (b) optical microscopy, (c) micro-computed tomography, and (d) scanning electron microscopy images of the moderate-density polyurethane-based foam. In all images, note the random cellular structure with open, connected pores and the moderate relative density of the matrix polymer material with void diameters in the tens to hundreds of micrometers. (e) Raw reference image in the digital image correlation sequence from an example compression experiment on the low-rate load frame described in \autoref{sec:exp} with typical approximate DIC area outlined by the dashed yellow box. (f) Similar to (e), an unprocessed reference image from an example tension experiment. }
	\label{fig:fig1_ex_foam}
\end{figure}

The rate-dependent nature of elastomeric foams is important in the design and use of impact mitigation materials (IMMs). As a staple IMM, elastomeric foams exist in many types, and new foams are an area of active development \citep{okolieocha2015,sun2018}, presenting modeling challenges. These randomly porous elastomeric polymers consist of a heterogeneous distribution of a viscoelastic polymer matrix material and gas-filled free space, as shown in the example images at different scales in \autoref{fig:fig1_ex_foam}(a)-(d). Depending on the matrix polymer and the geometrical properties of the pore space, these materials can exhibit high specific energy absorption due to dissipative micromechanical and matrix material effects and offer the capability to undergo recoverable deformations and hence repeated use \citep{mills2007polymer,tang2017}. Due to their porous nature, elastomeric foams are often lightweight with relative densities, $\rho/\rho_0$, where $\rho$ is the bulk density of the foam and $\rho_0$ is the bulk density of the matrix material, typically 0.3 and lower for impact mitigating foams. 
The overall stiffness of a foam is often tuned by adjusting the relative density during fabrication. Elastomeric foams have been widely adopted in part because they are inexpensively and straightforwardly fabricated at commercial process scales \citep{gibson97,mills2007polymer}. These aspects make elastic polymer foams attractive for modern protective equipment. However, to effectively design and deploy solutions using materials exhibiting complicated constitutive responses, material models must be sufficiently predictive and robust, requiring a thorough understanding of the material response. 

Experiments over a wide range of strain rates have been conducted to attempt to capture time-dependencies in the material response of elastomeric foams. Elevated-rate experiments can largely be grouped into three categories on the basis of strain rate. First, low-rate ``non-equilibrium'' experiments, in which the response displays non-negligible hysteresis, often due to viscous effects in the matrix, typically span strain-rates of $\dot \epsilon_{\rm applied} \approx 10^{-3}\,\text{s}^{-1}$ or lower to $10^{0}$\,s$^{-1}$ depending on the matrix material. Second, intermediate or impact-like rates, in which the response is hysteretic and other mechanisms, e.g., glassy transition of ligaments due to rate sensitivity \citep{yi2006} or microinertia \citep{calladine1984} may be active, occur over a typical strain-rate range of $\dot \epsilon_{\rm applied} \approx 10^{0}\,\text{s}^{-1}$ to $10^{3}\,\text{s}^{-1}$. Third are high or ballistic-like rates, in which shock-front propagation mechanics dominate the overall response \citep{xi2017,sun2016critshock}, macroscopic force equilibrium no longer exists, and internal gas viscosity may play a role \citep{mills_modelling_2002}. At high rates, the incident velocity, rather than an effective strain rate, is often used as the rate-defining metric since strain rate is highly nonuniform across a shock front. For sake of comparison, these rates roughly correspond to $\dot \epsilon_{\rm applied} \approx 10^{3}\,\text{s}^{-1}$ to $10^{7}\,\text{s}^{-1}$ and higher \citep{ma2009}. Here, we focus on the low- and intermediate-rate, non-equilibrium response of three densities of a highly rate-sensitive, polyurethane-based impact protection foam. This follows upon our study of the quasistatic response of this material, in which we described a hyperelastic constitutive model that captures the nonlinear stress--strain and lateral--axial strain response in compression and tension that remains predictive under multi-modal loading states \citep{landauer2019quasistatic}, and upon our extension of this model to capture finite-strain viscoelasticity of a closed-cell foam \citep{LI2022}.

The transition between quasistatic and low-rate non-equilibrium experiments for elastomeric foams can be difficult to define, since rate-independent dissipative mechanisms can exist. However, a natural distinction is whether viscous dissipative effects must be included to capture rate-sensitivity of, for example, the plateau stress or hysteresis. For elastic polymer foams, a broad selection of models in the literature include viscoelastic rate-sensitivity to capture dissipation as this is a common feature among polymers suited for impact protection and other foamed material applications \citep[e.g.,][]{yang2004,markert2008,anani2011,briody2012,chen2012,piero2012,ju2015,kossa2016,azizi2016,areias2017,esposito2020modeling,LI2022}. Although experimental treatments differ, universal testing machines are typically used for low-rate experiments and some intermediate-rate studies \citep[e.g.,][]{landauer2023dataframework, bhagavathula_density_2022}. Specialized fixturing can be used to probe a specific strain state \citep{liu2005,kossa2015,ling2018,bailly_strain_2020} or for model validation \citep{briody2012,LI2022}. 

For intermediate-rate experiments, a wide selection of equipment has been used to impose deformation, since universal testing systems struggle to perform at the necessary test conditions. Commonly, high-rate testing is achieved with a Kolsky bar\,---\,often specially modified for low-impedance materials\,---\,although other dedicated testing systems have been successfully employed \citep{markert2008,koumlis2019strain,bailly_strain_2020,landauer_stress_2020,bhagavathula_density_2022}. Impact-based systems are also relatively common, e.g., pneumatically driven \citep{markert2008}, flywheel-based \citep{viot2009}, or drop-weight systems \citep{roux2008,mills2009,mane2017,jeong2016,bailly_strain_2020,landauer_stress_2020,morrison2024temperature}. Small-strain rheological testing with time--temperature shifting can be used to predict linear viscoelastic parameters (viz. storage, $E'$, and loss, $E''$, moduli and the loss tangent, tan\,$\delta$) and has been used for polymer foams in both shear rheology and dynamic mechanical analysis (DMA) \citep{tang2017,sahraoui2000,landauer2023dataframework}. However, the effectiveness of these techniques is limited for large-deformation viscoelastic model development since the highly nonlinear response domain of these materials is inaccessible at the strain rates of interest, and the linear viscoelasticity assumptions necessary for time--temperature superposition and typical for the operational control loops in these instruments becomes inappropriate. High-speed servohydraulic load frames, one of the techniques used in this work, have also been used to good effect \citep{deshpande2001,rusch1969,shen2010,landauer2023dataframework}, although they are generally limited to the lower range of impact-like rates (ca. $\,\leq\,40\,\text{s}^{-1}$). 


Here, we discuss a suite of experiments on a blown-type, open-cell viscoelastic foam, and the development and specialization of a visco-hyperelastic continuum material model across three relative densities used in protective equipment, such as soccer padding or hockey helmets. The model is based on the framework of our previous quasistatic open-cell \citep{landauer2019quasistatic} and viscoelastic closed-cell \citep{LI2022} models for foamed elastomeric polymers. 
First, the homogeneous-deformation compression/tension experiments that inform constitutive model development as well as small-strain linear viscoelastic characterization of one density of the foam material as a function of rate and temperature are described (\autoref{sec:exp}). This is followed by the description of the constitutive model (\autoref{sec:model}), which employs a multi-mechanism rheological paradigm to capture the non-equilibrium material mechanics under large deformations. The model is then applied to all three foam densities using a viscoelastic generalization of the Equilibrium Constrained Genetic Algorithm (ECGA) framework of \citet{yan2026equilibrium} to estimate material parameters (\autoref{sec:cal}). With the calibrated model complete, validation experiments under an assortment of deformation modes and rates are shown to evaluate the model's predictive performance (\autoref{sec:result_validation}). Then, a new technique is introduced (\autoref{sec:loc}) to identify the onset of inhomogeneous deformation characteristics to distinguish between non-localizing and localizing strain fields, which places bounds on when homogeneous non-bifurcating models, i.e., models that do not capture sub-continuum buckling, are applicable. We close with a discussion of limitations (\autoref{sec:limitations}), a summary of the main contributions, and a discussion of future work (\autoref{conclusion}). 

\section{Materials and characterization experiments}\label{sec:exp}

This section describes the foam materials under consideration, specimen preparation, the homogeneous, finite-deformation experiments used for model calibration, and the small-strain, dynamic mechanical analysis experiments used to establish the temperature sensitivity of the foam material. 

\subsection{Materials} 
Three densities of a polyurethane-based, open-cell, viscoelastic impact protection foam (Poron XRD, Rogers Corp, Rogers, CT\footnote{Certain commercial equipment, instruments, software, or materials are identified in this paper in order to specify the experimental procedure adequately. Such identification is not intended to imply recommendation or endorsement by NIST, nor is it intended to imply that the materials or equipment identified are necessarily the best available for the purpose.}) were utilized: 095 series with a density of approximately 144\,kg/m$^3$, 125 series with a density of approximately 192\,kg/m$^3$, and 155 series with a density of approximately 240\,kg/m$^3$. These are referred to as the low, moderate, and high density foams, respectively, throughout. Images of the moderate-density foam at various scales are shown in \autoref{fig:fig1_ex_foam}(a)-(d). 

\subsection{Specimen preparation} 
Specimen preparation follows our previous work on characterization of the quasistatic response of these open-cell viscoelastic foams in \citet{landauer2019quasistatic}. To summarize, specimens were excised from as-received sheets of 12.9\,mm (approximately 0.5\,in) nominal thickness into cuboids of approximately 10\,mm $\times$ 10\,mm $\times$ 12.9\,mm for conducting compression experiments with a vertical bandsaw or into an ASTM D3574 dogbone geometry for conducting tensile experiments with a 75\,W CO$_2$ laser cutter. One face of each specimen was speckle painted for digital image correlation (DIC) analysis using an airbrush with white and black acrylic ink. Example images of the speckled specimens are shown in \autoref{fig:fig1_ex_foam}(e) and (f) for compression and tension specimens, respectively. Specimens for dynamic mechanical analysis (DMA) were excised from an as-received sheet using a graphite-lubricated brass hole-saw in a low-speed drill-press. To maintain force levels within the noise floor and the maximum force bounds of the DMA transducer, specimens of nominally 9\,mm diameter were used for the low- and moderate-density foams, and nominally 6\,mm diameter specimens were used for the high-density foam.

A number of experiments involving inhomogeneous deformation were carried out for validation testing of constitutive model predictions, as described in \autoref{sec:result_validation}. For indentation tests (\autoref{sec:indentation_v_poron}), rectangular specimens of about 25.4\,mm $\times$ 25.4\,mm $\times$ 12.9\,mm were laser cut and bonded to a rigid 6061 aluminum backer using cyanoacrylate-based instant adhesive. Double-lap shear specimens (\autoref{sec:shear}) were laser cut into rectangular prisms of about 30\,mm $\times$ 30\,mm $\times$ 12.9\,mm and bonded to aluminum backers that were custom-machined for a shear fixture. Tension specimens with circular holes (\autoref{sec:inhomo_tension_vp}) were laser cut from as-received sheets with the ASTM D3574 geometry with a pattern of holes on the surface. 

\subsection{Finite-deformation characterization experiments}
The homogeneous-deformation characterization experiments consist of a series of displacement-controlled triangular engineering strain histories in both compression and tension at increasing nominally constant engineering strain rates from $10^{-3}$\,s$^{-1}$ by decades to $10^{0}$\,s$^{-1}$. Each experiment was repeated three times on the same specimen with a wait time of 3 hours or longer between experiments, which was sufficiently long to allow relaxation of time-dependent material physics. To impose the desired strain ranges and rates, two experimental systems were employed. The first system, used for lower strain rates, is described in detail elsewhere \citep{landauer2018qdic,landauer2019quasistatic} and is used here with minimal reconfiguration. This system was designed to operate at strain-rate ranges up to about $10^{-1}$\,s$^{-1}$ for the typical tension and compression geometries described above, using a stepper-motor-driven linear actuator and a static-type load cell (222\,N nominal capacity, $\pm$0.15\,\% full scale repeatability and accuracy, upscale-downscale calibrated with calibrated reference masses). The expanded instrument measurement uncertainty associated with stress measurement\,---\,from the load cell calibration and areal measurement\,---\,was $\pm6.9$\,kPa. The system was operated at an elevated crosshead speed to match the nominal strain rate required. A high-resolution scientific camera with a tele-microscope lens was used with cross-polarized lighting to capture images for DIC analysis (see \autoref{tab:dic_table} for details). Two changes were made to this system to facilitate elevated-rate testing with little effect on the resultant data. First, the load cell was mounted underneath the specimen rather than in series with the actuator; and, second, a small dot of epoxy resin was used to affix the compression specimen to the bottom platen. Regarding the first, although linear elastic theory does not hold \citep{sun2018,ma2009}, it can provide an estimate of the validity of placing the load cell on the transmitted side of the specimen. The longitudinal wave speed is estimated as $C_{\rm P} = \sqrt{{E(1-\nu)}/{(\rho(1+\nu)(1-2\nu))}}$, where $E$ is the Young's modulus, $\nu$ is the Poisson's ratio, and $\rho$ is the density. Using the quasistatic parameters for the low-density foam as a worst-case and assuming an effective Poisson's ratio of 0.05 (again, informed by mean tangent Poisson's ratio from the low-density foam), we estimate $C_{\rm P} \approx 24.7$\,m/s. Therefore, the traveling wave reaches the load sensor in approximately $0.5\times10^{-4}$\,s, or about 0.0005\,\% of the total test time, and thus approximate force equilibrium holds. Regarding the second change, the epoxy spot size was typically less than 15\,\% of the bottom surface area of the otherwise well lubricated specimen. Although a small effect of frictional confinement was observed in the incremental displacement fields, this effect was similar to the low levels of friction observed in the lower-rate cases with no epoxy. 

\begin{figure}[!t]
	\centering{\includegraphics[width=0.85\textwidth]{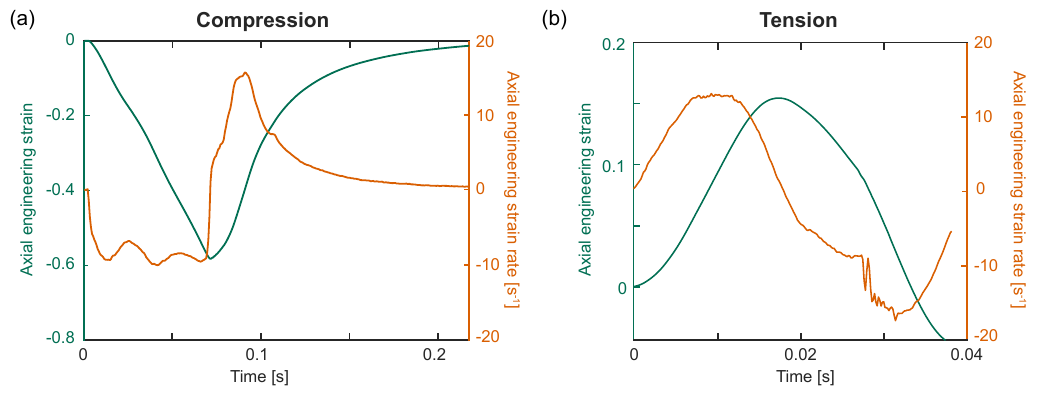}}
	\caption{Mean axial engineering strain and engineering strain rate as a function time for the moderate-rate load frame at a target strain rate of $10^1$\,s$^{-1}$. (a) For compression, the strain rate quickly rose and remained at approximately the target value; during unloading, the strain rate was governed by the fastest rate allowed by the material response. (b) For tension, a constant-acceleration region is followed by loading at the target rate and a partially controlled unloading.}
	\label{fig:strain_strainrate_plots}
\end{figure}

The second system was designed for intermediate-rate experiments using a linear induction motor and was originally described in \citet{landauer2019dissertation} and \citet{LI2022}. Since the foams considered in this work are considerably softer than the closed-cell foam of \citet{LI2022}, a load transducer with lower capacity and finer resolution (LCFD 10, Omega Engineering, Stamford, CT) was utilized for force data collection (44.5\,N nominal capacity, 0.15\,\% full scale range uncertainty, upscale-downscale calibrated with calibrated reference masses). Instrument measurement uncertainty associated with stress measurement was $\pm1.4$\,kPa. The same high-speed camera and optics as in the first system were used for imaging and DIC analysis (see \autoref{tab:dic_table} for details). At overlapping nominal strain rates, the performance of the two systems was equivalent \citep{LI2022}. For the highest strain rate, machine performance was less ideal than the lower-rate cases due to, for example, inertial effects and the feedback-controlled nature of the induction motor, particularly in tension-type experiments. As an illustration of this limitation for a target engineering strain rate of $10^1$\,s$^{-1}$, \autoref{fig:strain_strainrate_plots} shows the mean axial engineering strain magnitudes and engineering strain rates as a function of time in both compression and tension. For these experiments, the axial engineering strain histories were measured from the mean of the DIC strain fields, and the strain rate histories were computed directly from the strain histories via central differencing. Due to the large discrepancy between the target triangular engineering strain history and the measured strain history at this rate, the characterization experiments included in this work were limited to a maximum strain rate of $10^0$\,s$^{-1}$, for which the instrument remained robust. 

\subsection{Dynamic mechanical analysis} 
\begin{figure}[!t]
	\centering{\includegraphics[width=1\textwidth]{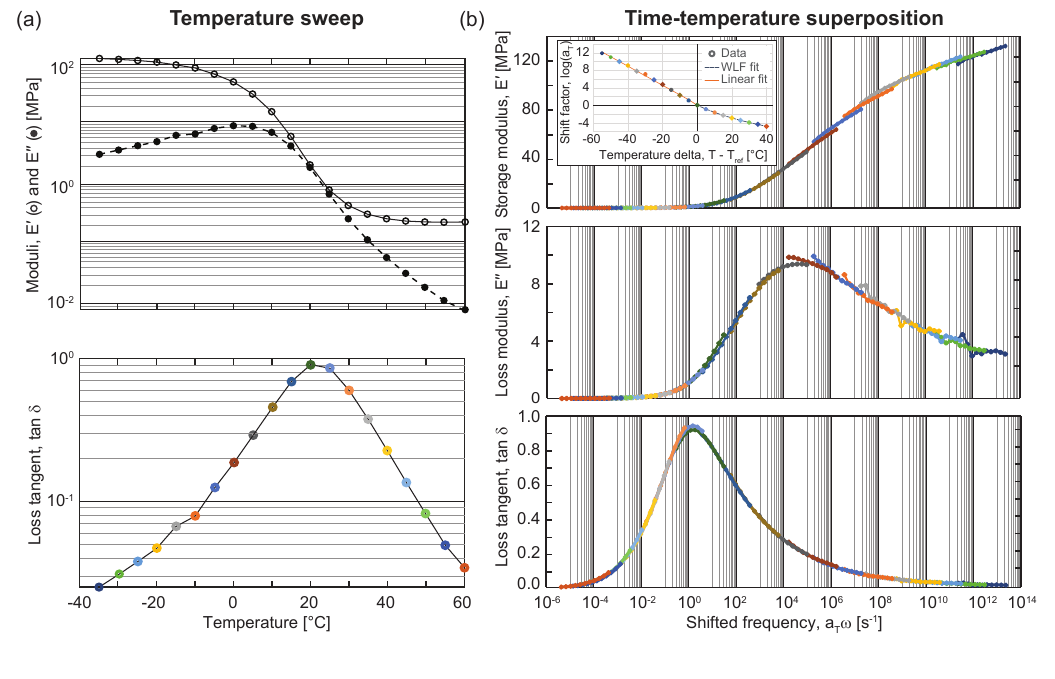}}
	\caption[Example DMA data for the moderate density foam]{Example small-amplitude oscillatory strain data for the moderate-density foam. (a) A temperature sweep at 10\,Hz demonstrating the thermal sensitivity of the storage modulus $E'$, loss modulus $E''$, and loss tangent $\tan\delta$, showing a broad transition between the glassy and rubbery states with an approximate glass transition temperature of $T_{\rm g} \approx 21$\,\textdegree C. Note that a logarithmic vertical axis is used to highlight the brief crossing point of the storage and loss moduli. (b) Time-temperature superposition using small-strain linear viscoelastic assumptions over the temperatures and frequencies tested at a reference temperature of $20$\,\textdegree C. The test temperatures for each segment are color coded according the colors in the loss tangent temperature sweep plot in (a). The inset plot shows the shift factors used to generate these composite curves with a linear fit for sub-reference and WLF fit for super-reference temperatures, which indicate that linear viscoelastic assumptions are reasonable. The standard uncertainties were estimated to be $\pm$\,5\,\% on frequency, $\pm$\,8\,\% on moduli, and $\pm$\,0.1\,\textdegree C on temperature measurements.}
	\label{fig:dma_data}
\end{figure}

Dynamic mechanical analysis (DMA) was conducted following a procedure similar to that described in \citet{landauer2023dataframework}. To begin, specimens were epoxied to the DMA platen with a high-temperature, rigid, two-part resin, and preliminary experiments were conducted to ascertain the linear viscoelastic regime and the relaxation and thermal equilibration times of each foam. Then, frequency sweep experiments from approximately 0.048\,Hz to 4.8\,Hz were conducted at set-point temperatures from 60\,\textdegree C to -35\,\textdegree C in steps of 5\,\textdegree C with thermal resets to eliminate history effects in the polymer of 1 h or more between each temperature, and thus results are independent temperature sweep direction. At each step, a compressive pre-strain of approximately \,7\,\% or pre-load of approximately 0.3\,N (whichever was smaller) was applied. Care was taken to remain in the linear viscoelastic regime. Representative data for thermal sensitivity and time-temperature superposition under small-strain linear viscoelastic assumptions for the DMA experiments are shown for the moderate-density foam in \autoref{fig:dma_data}. Shifting was conducted via the Williams-Landel-Ferry (WLF) equation \cite{williams_temperature_1955} with horizontal shift factors $a_{\rm T}$ manually determined to concurrently optimize moduli and loss tangent continuity (shift factors are shown in the inset plot in \autoref{fig:dma_data}b). All other experiments (tension and compression and validation tests) were conducted at room temperature, approximately 23\,\textdegree C, which is close the glass transition temperature inferred from the DMA results ($T_{\rm g} \approx 21$\,\textdegree C) and is a relatively temperature-sensitive domain for these foams, potentially contributing to scatter in the data if small changes in ambient temperature occurred.

\section{Constitutive model}\label{sec:model}
The finite-deformation viscoelastic constitutive model described in this section is based on the framework introduced in \citet{LI2022} with specializations, modifications, and simplifications to capture the rate-dependent behavior of the open-cell viscoelastic foams of \autoref{sec:exp}. 
Finite deformations are described by the motion function $\bfy = \bfchi(\bfx, t)$ that maps material points $\bfx$ to points $\bfy$ in the deformed configuration at each time $t$.\footnote{{\it Notation:} The symbols $\nabla$ and Div denote the gradient and divergence with respect to the material point $\bfx$, while grad and div denote these operators with respect to the point $\bfy = \bfchi(\bfx, t)$. The quantities $\Tr\bfA$, $\det\bfA$, $\dev\bfA$, $\sym\bfA$, and $\skw\bfA$ denote the trace, determinant, deviatoric part, symmetric part, and skew part of a tensor $\bfA$, respectively. The symbol $\otimes$ denotes the tensor product. } The deformation gradient is $\bfF = \nabla\bfchi$, such that the volume ratio is strictly positive, i.e., $J = \Det\bfF > 0$, and the displacement field is $\bfu(\bfx,t) = \bfchi(\bfx,t)-\bfx$. The left polar decomposition of $\bfF$ is $\bfF = \bfV\bfR$, where $\bfV$ is the left stretch tensor and $\bfR$ is the rotation tensor. The tensor $\bfV$ admits the spectral decomposition $\bfV = \sum_{i=1}^3(\lambda_i)\bfl_i\otimes\bfl_i$, where $\{\bfl_i|i=1,2,3\}$ are the principal directions of $\bfV$ and $\{\lambda_i|i=1,2,3\}$ are the principal stretches. Finally, the spatial logarithmic (Hencky) finite-strain tensor is $\bfE = \sum_{i=1}^3(\ln\lambda_i)\bfl_i\otimes\bfl_i$. 

\begin{figure}[!t]
	\centering
\includegraphics[width=4.5in]{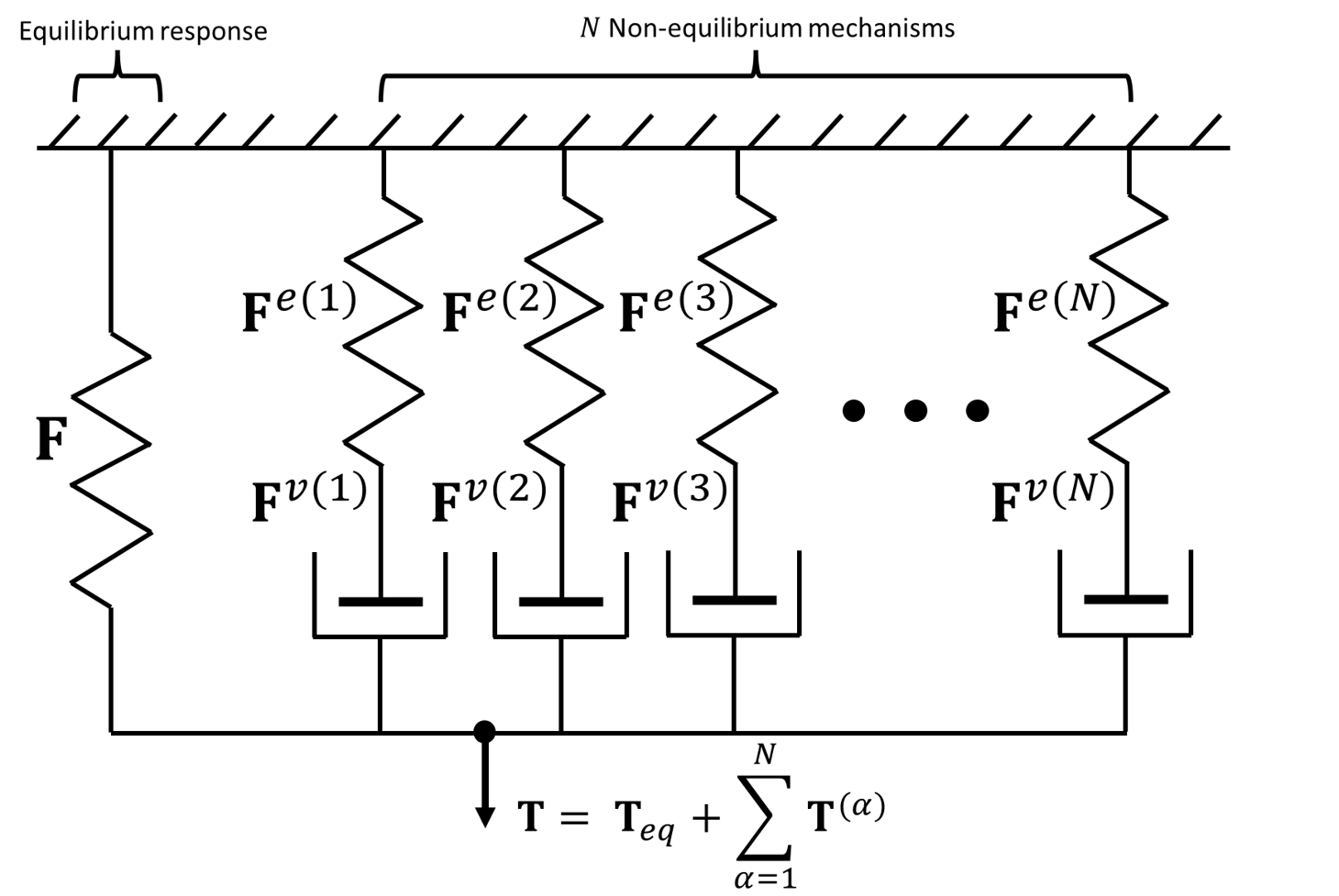}
\caption{A rheological schematic of the nonlinear, finite-deformation, viscoelastic constitutive model consisting of one equilibrium mechanism in parallel with $N$ non-equilibrium mechanisms, which are described by nonlinear, Maxwell-like elements.}\label{rheo}
\end{figure}

The constitutive model utilizes a decomposition of the mechanical response into an equilibrium hyperelastic contribution and a number of dissipative, non-equilibrium contributions, as illustrated rheologically in \autoref{rheo}. A key idealization of the nonlinear viscoelastic constitutive model is isotropy, since this is a good approximation for the elastomeric foams under consideration here, as shown in \citep{landauer2019quasistatic}. The equilibrium response is described by the hyperelastic model of \citet{landauer2019quasistatic}, which is summarized in \autoref{sec:eq0}, while the non-equilibrium response consists of $N$ nonlinear, Maxwell-like mechanisms in parallel with the equilibrium response. The mathematical structure of the non-equilibrium response is slightly modified compared to the constitutive equations of \citet{LI2022} and is described in \autoref{sec:noneq0}. 

\subsection{Equilibrium response}\label{sec:eq0}
To specify the equilibrium hyperelastic response, the model utilizes the free-energy density function of our prior work \citep{landauer2019quasistatic}, which depends on the set of invariants of the logarithmic strain tensor $\bfE$ proposed by \citet{criscione2000} and denoted as $\{K_1, K_2, K_3\}$. The first invariant $K_1 = \Tr(\bfE) = \ln(J)$ corresponds to the logarithmic volumetric strain. The second invariant $K_2 = |\dev(\bfE)| \ge 0$ represents the magnitude of distortion independent of volume change. The third invariant $K_3 = 3\sqrt{6}\,\Det(\bfN) \in [-1, 1]$, where $\bfN = \dev(\bfE)/K_2$ is the tensorial direction of deviatoric strain, represents the mode of constant-volume distortion. For example, $K_3 = -1$ in simple compression, $K_3 = 0$ in simple shear, and $K_3 = 1$ in simple tension. 

As in \citet{landauer2019quasistatic}, the equilibrium free-energy density function $\tilde\psi_{\rm eq}(K_1,K_2,K_3)$ is given by the following form: 
\begin{equation}
    \tilde\psi_{\rm eq}(K_1,K_2,K_3) = G_0 \left[ X_{\rm eq}(K_1) K_2^2 + L_{\rm eq}(K_2, K_3) \right] + B f_{\rm eq}(K_1),
    \label{eq:free_energy}
\end{equation}
where $G_0$ and $B$ are the ground-state, equilibrium shear and bulk moduli, respectively, and $X_{\rm eq}(K_1)$, $L_{\rm eq}(K_2,K_3)$, and $f_{\rm eq}(K_1)$ are phenomenological functions. The $G_0 X_{\rm eq}(K_1) K_2^2$ term in \autoref{eq:free_energy} describes the coupling between the volumetric and distortional responses. For simplicity, the dependence on $K_2$ in this term is only of second order, so that the function $X_{\rm eq}(K_1)$ characterizes the volumetric-strain-dependence of the instantaneous shear modulus. Then, the $G_0 L_{\rm eq}(K_2, K_3)$ term in \autoref{eq:free_energy} characterizes the higher-order distortional response, which is assumed to be independent of the volumetric strain. Finally, the $B f_{\rm eq}(K_1)$ term in \autoref{eq:free_energy} characterizes the purely volumetric response. 

The equilibrium contribution to the Cauchy stress, $\bfT_{\rm eq}$, then follows from the free-energy density function \autoref{eq:free_energy} as
\begin{equation}\label{eq:cauchy2}
\begin{split}
\bfT_{\rm eq} &= J^{-1}\left[\dfrac{\partial \tilde\psi_{\rm eq}}{\partial K_1}\id + \dfrac{\partial\tilde\psi_{\rm eq}}{\partial K_2}\bfN +\dfrac{\partial\tilde\psi_{\rm eq}}{\partial K_3}\dfrac{1}{K_2}\bfY\right]\\[4pt]
&= J^{-1}\left[\left(G_0\dfrac{dX_{\rm eq}}{dK_1}K_2^2 + B\dfrac{d f_{\rm eq}}{dK_1}\right)\id + G_0\left(2X_{\rm eq}(K_1)K_2 + \dfrac{\partial L_{\rm eq}}{\partial K_2}\right)\bfN + G_0\dfrac{\partial L_{\rm eq}}{\partial K_3}\dfrac{1}{K_2}\bfY\right],
\end{split}
\end{equation}
where $\id$ is the identity tensor and $\bfY = 3\sqrt{6}\bfN^2 - \sqrt{6}\id - 3K_3\bfN$.

Specific forms for the phenomenological fitting functions $X_{\rm eq}(K_1)$, $L_{\rm eq}(K_2,K_3)$, and $f_{\rm eq}(K_1)$ may then be selected to specialize the model for different foam materials. Here, we utilize the functional forms of \citet{landauer2019quasistatic}, which are summarized below.
\begin{enumerate}

\item The distortional/volumetric-coupling response function, $X_{\rm eq}$:

\begin{equation}
    X_{\rm eq}( K_1 ) = \dfrac{1}{2} \left( X_1^\prime + X_2^\prime \right) K_1 + \dfrac{\Delta_{\rm K}}{2} \left( X_1^\prime - X_2^\prime \right) \ln \left[\dfrac{ \cosh \left( (K_1 - K_1^0)/\Delta_{\rm K} \right) }{\cosh \left(K_1^0/\Delta_{\rm K} \right)}\right] + 1.
    \label{eq:coupled_fit}
\end{equation}
The function $X_{\rm eq}(K_1)$ has a close-to-linear dependence on $K_1$, while accounting for different slopes before and after the transition to the plateau regime in the stress--strain response in compression. The parameter $X_1^\prime$ describes the slope before the transition to the plateau regime, and $X_2^\prime$ describes the slope after the transition to the plateau regime. The parameter $K_1^0 < 0$ represents the value of $K_1$ at which the transition to the plateau regime takes place, and the parameter $\Delta_{\rm K} > 0 $ denotes the range of $K_1$ over which the transition to the plateau regime occurs.

\item The higher-order distortional response function, $L_{\rm eq}$:
\begin{equation}
    L_{\rm eq} (K_2,K_3) = C_0 K_{2}^{p} + C_1 \left( 1 + K_3 \right) K_2^{q}.
    \label{eq:L_1}
\end{equation}
The function $L_{\rm eq}$ depends on both $K_2$ and $K_3$ and captures the higher-order distortional response. In simple compression, $K_3=-1$, and the second term in \autoref{eq:L_1} vanishes, enabling tension-compression asymmetry to be captured by the model beyond that described by the purely volumetric response function $f_{\rm eq}$. In \autoref{eq:L_1}, $C_0 \geq 0 $ and $C_1 \geq 0$ are constant parameters paired with $p > 2$ and $q > 2$, where $p$ and $q$ are the higher-order exponents.

\item The purely volumetric response function, $f_{\rm eq}$:
\begin{equation}\label{eq:f_vr}
\dfrac{df_{\rm eq}}{dK_1} = \dfrac{J^{C_2}-1}{C_2} + C_3J\left[J^{-r} - \left(\dfrac{1-J_{\rm min}}{J-J_{\rm min}}\right)^r\right],
\end{equation}
%
%
where $J = \exp(K_1)$. The function $f_{\rm eq}$ captures the response to purely volumetric deformation and is written above in terms of its derivative since only $d f_{\rm eq}/dK_1$ appears in the expression for the equilibrium Cauchy stress \autoref{eq:cauchy2}. The first term, parameterized by $C_2$ in \autoref{eq:f_vr}, is responsible for capturing the volumetric response in tension and in compression up to the transition to the plateau regime. The second term, parameterized by $\{J_{\rm min},r,C_3\}$, captures the stiffening behavior due to volume change in the densification regime during compression. The parameter $0<J_{\rm min}<1$ sets the minimum allowable value of $J$, at which $d f_{\rm eq}/dK_1$ diverges as $J$ approaches $J_{\rm min}$ from above, and the parameter $r$ controls the nature of this divergence. 
\end{enumerate}

\subsection{Non-equilibrium response}\label{sec:noneq0}
The viscoelastic material response is described by $N$ nonlinear, Maxwell-like, non-equilibrium mechanisms in parallel with the equilibrium response, as illustrated rheologically in \autoref{rheo}. In each non-equilibrium mechanism with index denoted by $\alpha$, the deformation gradient $\bfF$ is multiplicatively decomposed into elastic and viscous parts: $\bfF = \bfF^{{\rm e}(\alpha)}\bfF^{{\rm v}(\alpha)}$, where $\bfF^{{\rm e}(\alpha)}$ with $ J^{{\rm e}(\alpha)} = \Det{\bfF^{{\rm e}(\alpha)}} >0$ is the non-equilibrium elastic distortion, which corresponds to the deformation of spring $\alpha$, and $\bfF^{{\rm v}(\alpha)}$ with $ J^{{\rm v}(\alpha)} = \Det{\bfF^{{\rm v}(\alpha)}} >0$ is the viscous distortion corresponding to the deformation of dashpot $\alpha$. 

\subsubsection{Non-equilibrium free-energy function} 
The contribution to the free-energy density due to each non-equilibrium spring $\alpha$ in \autoref{rheo} is based on the respective elastic logarithmic strain tensor $\bfE^{{\rm e}(\alpha)}$ for mechanism $\alpha$, which follows from the following definitions: $\bfF^{{\rm e}(\alpha)} = \bfV^{{\rm e}(\alpha)}\bfR^{{\rm e}(\alpha)}$, the left polar decomposition of $\bfF^{{\rm e}(\alpha)}$; $\bfV^{{\rm e}(\alpha)} = \sum_{i=1}^3(\lambda_i^{{\rm e}(\alpha)})\bfl_i^{{\rm e}(\alpha)}\otimes\bfl_i^{{\rm e}(\alpha)}$, the spectral decomposition of $\bfV^{{\rm e}(\alpha)}$; and $\bfE^{{\rm e}(\alpha)}=\sum_{i=1}^3(\ln\lambda_i^{{\rm e}(\alpha)})\bfl_i^{{\rm e}(\alpha)}\otimes\bfl_i^{{\rm e}(\alpha)}$, the elastic logarithmic strain tensor. As in the formulation of the equilibrium free-energy density, we utilize an analogous set of invariants based on the elastic logarithmic finite-strain tensor $\bfE^{{\rm e}(\alpha)}$\,---\,namely, $K_1^{{\rm e}(\alpha)} = \Tr(\bfE^{{\rm e}(\alpha)})$, $K_2^{{\rm e}(\alpha)} = |\dev(\bfE^{{\rm e}(\alpha)})| \ge 0$, and $K_3^{{\rm e}(\alpha)}= 3\sqrt{6}\,\Det(\bfN^{{\rm e}(\alpha)}) \in [-1, 1]$, where $\bfN^{{\rm e}(\alpha)} = \dev(\bfE^{{\rm e}(\alpha)})/K_2^{{\rm e}(\alpha)}$. Then, the non-equilibrium free-energy density function $ \tilde\psi_{\rm neq}^{(\alpha)}(K_1^{{\rm e}(\alpha)},K_2^{{\rm e}(\alpha)},K_3^{{\rm e}(\alpha)})$ for each mechanism $\alpha$ is taken in a functional form analogous to \autoref{eq:free_energy}: 
\begin{equation}\label{eq:non_eq_free_energy}
    \tilde\psi_{\rm neq}^{(\alpha)}(K_1^{{\rm e}(\alpha)},K_2^{{\rm e}(\alpha)},K_3^{{\rm e}(\alpha)}) = G_{\rm neq}^{(\alpha)} \left[ X^{(\alpha)}_{\rm neq}(K_1^{{\rm e}(\alpha)}) (K_2^{{\rm e}(\alpha)})^2 + L_{\rm neq}^{(\alpha)}(K_2^{{\rm e}(\alpha)}, K_3^{{\rm e}(\alpha)}) \right] + B_{\rm neq}^{(\alpha)} f^{(\alpha)}_{\rm neq}(K_1^{{\rm e}(\alpha)}),
\end{equation}
where $G_{\rm neq}^{(\alpha)}$ and $B_{\rm neq}^{(\alpha)}$ are the ground-state, non-equilibrium shear and bulk moduli for each mechanism $\alpha$, respectively, and $X^{(\alpha)}_{\rm neq}(K_1^{{\rm e}(\alpha)})$, $L^{(\alpha)}_{\rm neq}(K_2^{{\rm e}(\alpha)}, K_3^{{\rm e}(\alpha)})$, and $f^{(\alpha)}_{\rm neq}(K_1^{{\rm e}(\alpha)})$ are phenomenological functions. 

The contribution to the Cauchy stress due to each non-equilibrium mechanism $\alpha$ then follows from \autoref{eq:non_eq_free_energy} as
\begin{equation}
 \begin{aligned}
\bfT^{(\alpha)} &= J^{{\rm e}(\alpha)-1}\left[\dfrac{\partial \tilde\psi_{\rm neq}^{(\alpha)}}{\partial K_1^{{\rm e}(\alpha)}}\id + \dfrac{\partial\tilde\psi_{\rm neq}^{(\alpha)}}{\partial K_2^{{\rm e}(\alpha)}}\bfN^{{\rm e}(\alpha)} +\dfrac{\partial\tilde\psi_{\rm neq}^{(\alpha)}}{\partial K_3^{{\rm e}(\alpha)}}\dfrac{1}{K_2^{{\rm e}(\alpha)}}\bfY^{{\rm e}(\alpha)}\right]\\[4pt]
& = J^{{\rm e}(\alpha)-1} \Biggl[\left(G_{\rm neq}^{(\alpha)}\dfrac{dX^{(\alpha)}_{\rm neq}}{dK_1^{{\rm e}(\alpha)}}(K_2^{{\rm e}(\alpha)})^2 + B_{\rm neq}^{(\alpha)}\dfrac{df^{(\alpha)}_{\rm neq}}{dK_1^{{\rm e}(\alpha)}}\right)\id\\[4pt]
&\qquad\qquad\qquad+ G_{\rm neq}^{(\alpha)}\left(2X^{(\alpha)}_{\rm neq}(K_1^{{\rm e}(\alpha)})K_2^{{\rm e}(\alpha)} + \dfrac{\partial L^{(\alpha)}_{\rm neq}}{\partial K_2^{{\rm e}(\alpha)}}\right)\bfN^{{\rm e}(\alpha)} + G_{\rm neq}^{(\alpha)}\dfrac{\partial L^{(\alpha)}_{\rm neq}}{\partial K_3^{{\rm e}(\alpha)}}\dfrac{1}{K_2^{{\rm e}(\alpha)}}\bfY^{{\rm e}(\alpha)}\Biggr],
 \end{aligned}
\label{eq:cauchy_neq2}
\end{equation}
where $\bfY^{{\rm e}(\alpha)} = 3\sqrt{6}{\bfN^{{\rm e}(\alpha)}}^2 - \sqrt{6}\id - 3K_3^{{\rm e}(\alpha)}\bfN^{{\rm e}(\alpha)}$. The total Cauchy stress is $\bfT = \bfT_{\rm eq} + \sum_{\alpha=1}^N\bfT^{(\alpha)}$. 

To reduce model complexity, for all non-equilibrium branches $\alpha$, the functional forms adopted for $X_{\rm neq}^{(\alpha)}$, $L_{\rm neq}^{(\alpha)}$, and $f_{\rm neq}^{(\alpha)}$ are the same as those adopted for $X_{\rm eq}$, $L_{\rm eq}$, and $f_{\rm eq}$ in \autoref{sec:eq0}. Moreover, the parameters in $X_{\rm neq}^{(\alpha)}$ and $f_{\rm neq}^{(\alpha)}$ are taken to be the same as those in $X_{\rm eq}$ (\autoref{eq:coupled_fit}) and $f_{\rm eq}$ (\autoref{eq:f_vr}). In the function $L_{\rm neq}^{(\alpha)}$, as in \citet{LI2022}, we allow for several branch-dependent parameters to capture viscoelastic behavior in tension and take 
\begin{equation}
    L_{\rm neq}^{(\alpha)} (K_2^{{\rm e}(\alpha)},K_3^{{\rm e}(\alpha)}) = C_{0} (K_{2}^{{\rm e}(\alpha)})^{p} + C_{1{\rm neq}}^{(\alpha)} \left( 1 + K_3^{{\rm e}(\alpha)} \right) (K_2^{{\rm e}(\alpha)})^{q_{\rm neq}^{(\alpha)}},
    \label{eq:L_1_neq}
\end{equation}
where the parameters $C_0$ and $p$ are branch-independent and the same as those in $L_{\rm eq}$ (\autoref{eq:L_1}), and the parameters $C_{1{\rm neq}}^{(\alpha)}\ge 0$ and $q_{\rm neq}^{(\alpha)}>2$ are branch-dependent. Finally, as in \citet{LI2022}, the ratio of the ground-state, non-equilibrium bulk and shear moduli for each non-equilibrium branch $\alpha$ is set to be the same as the ratio in the equilibrium branch, i.e., $B_{\rm neq}^{(\alpha)} / G_{\rm neq}^{(\alpha)} = B / G_{0}$ for all non-equilibrium mechanisms $\alpha$. In this way, the number of tunable parameters for the elastic response of each non-equilibrium branch $\alpha$ is reduced to three: $G_{\rm neq}^{(\alpha)}$, $C_{1{\rm neq}}^{(\alpha)}$, and $q_{\rm neq}^{(\alpha)}$. 

\subsubsection{Flow rule}\label{sec:flowrule}
The evolution of $\bfF^{{\rm v}(\alpha)}$ due to the response of the dashpot associated with each non-equilibrium mechanism $\alpha$ in \autoref{rheo} is
\begin{equation}\label{eq:evolution}
 \begin{aligned}
\dot{\bf F}^{{\rm v}(\alpha)} = {\bf D}^{{\rm v}(\alpha)}{\bf F}^{{\rm v}(\alpha)},\quad{\bf F}^{{\rm v}(\alpha)}({\bf x},t=0)=\id,
\end{aligned}
\end{equation}
where $\bfD^{{\rm v}(\alpha)}$ is the symmetric viscous stretching tensor for non-equilibrium mechanism $\alpha$, and we have assumed that the viscous spin tensor for each mechanism is zero, i.e., $\bfW^{{\rm v}(\alpha)} = \skw(\dot\bfF^{{\rm v}(\alpha)}\bfF^{{\rm v}(\alpha)-1}) = \zed$ for each $\alpha$. The flow rule for each mechanism $\alpha$ is the constitutive equation relating the viscous stretching tensor $\bfD^{{\rm v}(\alpha)}$ to the Mandel stress tensor for that mechanism $\bfM^{{\rm e}(\alpha)} = J^{{\rm e}(\alpha)}\bfR^{{\rm e}(\alpha)}{}^\top\bfT^{(\alpha)} \bfR^{{\rm e}(\alpha)}$, which we take to be 
\begin{equation}\label{eq:Dv}
\bfD^{{\rm v}(\alpha)} = \dfrac{\dev(\bfM^{{\rm e}(\alpha)})}{2\eta^{(\alpha)}} + \dfrac{\Tr(\bfM^{{\rm e}(\alpha)})}{9\kappa^{(\alpha)}}\id,
\end{equation}
where $\eta^{(\alpha)}$ and $\kappa^{(\alpha)}$ are the shear and bulk viscosities, respectively, for mechanism $\alpha$. 

In \autoref{eq:Dv}, there are two key departures from our previous work \citep{LI2022}. 
\begin{enumerate}
    \item First, \autoref{eq:Dv} assumes Newtonian rheological behavior. In \citet{LI2022}, the structure of the flow rule allowed for non-Newtonian rheological behavior, in which the viscosities $\eta^{(\alpha)}$ and $\kappa^{(\alpha)}$ could depend on scalar invariants of the viscous stretching tensor $\bfD^{{\rm v}(\alpha)}$ in a power-law form, and one non-Newtonian, shear-thinning non-equilibrium branch was used in tandem with several Newtonian branches to capture the nonlinear viscoelastic behavior of a closed-cell elastomeric foam. Here, we find that it is only necessary to invoke Newtonian non-equilibrium branches to capture the responses of the open-cell viscoelastic foams under consideration, reducing the flow rule to the form given in \autoref{eq:Dv}. As in \citet{LI2022}, we simplify the model by taking the shear and bulk viscosities to be the same, i.e., $\eta^{(\alpha)} = \kappa^{(\alpha)}$ for each mechanism. 
    \item Second, we introduce viscous-volumetric-strain dependence to the viscosities through the quantity $K_{1}^{{\rm v}(\alpha)}$, which is defined as $K_{1}^{{\rm v}(\alpha)} = \ln{J^{{\rm v}(\alpha)}} = \ln{J/J^{{\rm e}(\alpha)}} = \ln{J} - K_1^{{\rm e}(\alpha)}$. This allows the model to account for dependence of the viscosities on the magnitude of the viscous volumetric strain in compression, which gives rise to increased hysteresis when the material is deformed into the densification regime in compression. This effect is captured through the phenomenological dependence of $\eta^{(\alpha)}$ and $\kappa^{(\alpha)}$ on $K_{1}^{{\rm v}(\alpha)}$ in the following form: 
    \begin{equation}\label{eq:eta_eq}
       \eta^{(\alpha)}(K_{1}^{{\rm v}(\alpha)}) = \kappa^{(\alpha)}(K_{1}^{{\rm v}(\alpha)}) = \eta^{(\alpha)}_0\left\{
       \begin{array}{@{}ll@{}}
          1 & \text{if}\ K_{1}^{{\rm v}(\alpha)}>K_{10}^{\rm v}, \\[4pt]
          1 + C_4(K_{10}^{\rm v} - K_{1}^{{\rm v}(\alpha)})^{C_{\rm e}} & \text{if}\ K_{1}^{{\rm v}(\alpha)} \leq K_{10}^{\rm v},
       \end{array}\right.
    \end{equation} 
    where $\eta^{(\alpha)}_0$ is the branch-dependent baseline viscosity when the viscous volumetric strain $K_{1}^{{\rm v}(\alpha)}$ is zero and $\{C_4,C_{\rm e},K^{\rm v}_{10}\}$ is a set of branch-independent parameters aimed at capturing the dependence of the viscosity on $K_{1}^{{\rm v}(\alpha)}$. 
    The material parameter $K_{10}^{\rm v} < 0$ determines the threshold value of $K_{1}^{{\rm v}(\alpha)}$, at which the viscous-volumetric-strain-dependence activates in densification, and the shear and bulk viscosities are constant in a given branch $\alpha$ if $K_{1}^{{\rm v}(\alpha)} > K_{10}^{\rm v}$. The constant material parameter $C_4 > 0$ and exponent $C_{\rm e} > 0$ control the viscous-volumetric-strain-dependence for $K_{1}^{{\rm v}(\alpha)} < K_{10}^{\rm v}$ and describe a monotonically increasing viscosity as $K_{1}^{{\rm v}(\alpha)}$ decreases beyond $K_{10}^{\rm v}$. Since the viscous volumetric strain $K_{1}^{{\rm v}(\alpha)}$ is greater than zero in tension and less than zero in compression, \autoref{eq:eta_eq} does not affect the viscosities in tension or during the initial stages of compression but introduces an increase in the shear and bulk viscosities during the densification phase of compression. \autoref{eq:eta_eq} is inspired by features of the viscoelastic constitutive equations in \citet{bergstrom1998constitutive}, namely, the dependence of the rheological behavior on invariants of the viscous distortion. 
\end{enumerate}
Because the shear and bulk viscosities are equal, a characteristic relaxation time for each branch $\alpha$ may be defined as $t_0^{(\alpha)} = \eta_0^{(\alpha)}/G_{\rm neq}^{(\alpha)}$. Then, summarizing the parameters associated with the non-equilibrium response, each non-equilibrium mechanism $\alpha$ involves four branch-dependent parameters $\{t_0^{(\alpha)},G^{(\alpha)}_{\rm neq},C^{(\alpha)}_{1{\rm neq}},q^{(\alpha)}_{\rm neq}\}$. In addition, we reiterate that the parameters $K_{10}^{\rm v}$, $C_4$ and $C_{\rm e}$ are branch independent and constrained to be the same across all the non-equilibrium branches. 

\section{Application to open-cell viscoelastic foams}\label{sec:cal}

\begin{figure}[!t]
\centering
	\includegraphics[width=0.99\textwidth]{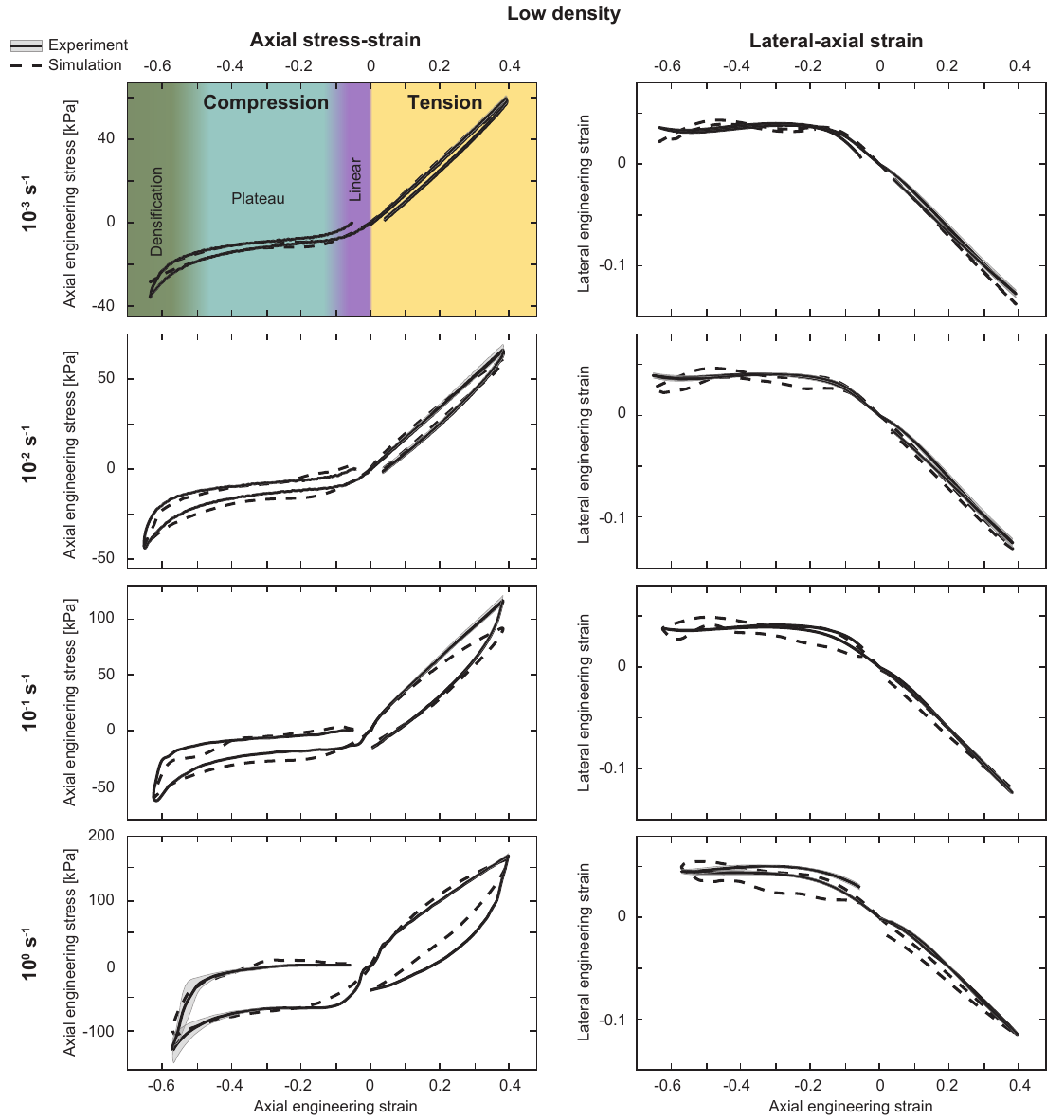}  
	\caption{Tension-compression response over four decades of strain rate in displacement-controlled load-unload experiments with corresponding model fits. The \textit{Upper Left} plot includes highlighted regions of the four deformation regimes (viz. tension-linear-plateau-densification) typical of these open-cell viscoelastic foams. (\textit{Left}) Axial engineering stress versus axial engineering strain curves and (\textit{Right}) lateral engineering strain versus axial engineering strain curves for the open-cell viscoelastic foam of low density (144\,kg/m$^3$) over the strain-rate range $10^{-3}$\,s$^{-1}$ to $10^0$\,s$^{-1}$. The solid lines with shaded error areas (one standard deviation, three experiments) indicate experimental data, and the dashed lines are constitutive model fits. To best show the characteristics of the curves, the stress axes differ in range for each strain rate.}
	\label{fig:poron95_v_simple}
\end{figure}

\begin{figure}[!t]
\centering
	\includegraphics[width=0.99\textwidth]{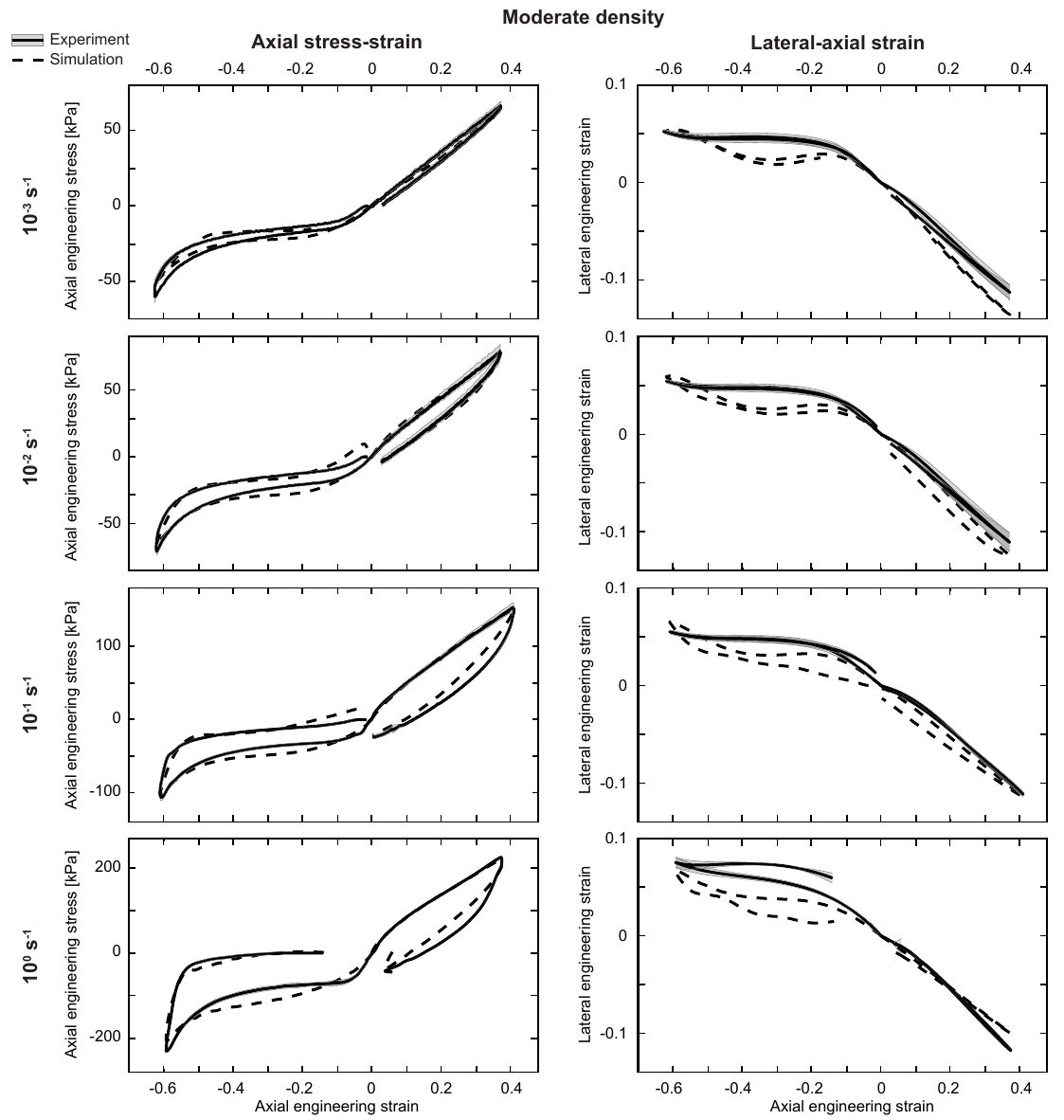}  
	\caption{(\textit{Left}) Axial engineering stress versus axial engineering strain curves and (\textit{Right}) lateral engineering strain versus axial engineering strain curves for the open-cell viscoelastic foam of moderate density (192\,kg/m$^3$) over the strain-rate range $10^{-3}$\,s$^{-1}$ to $10^0$\,s$^{-1}$. The solid lines with shaded error areas (one standard deviation, three experiments) indicate experimental data, and the dashed lines are constitutive model fits. To best show the characteristics of the curves, the stress axes differ in range for each strain rate.}
	\label{fig:poron125_v_simple}
\end{figure}

\begin{figure}[!t]
\centering
	\includegraphics[width=0.99\textwidth]{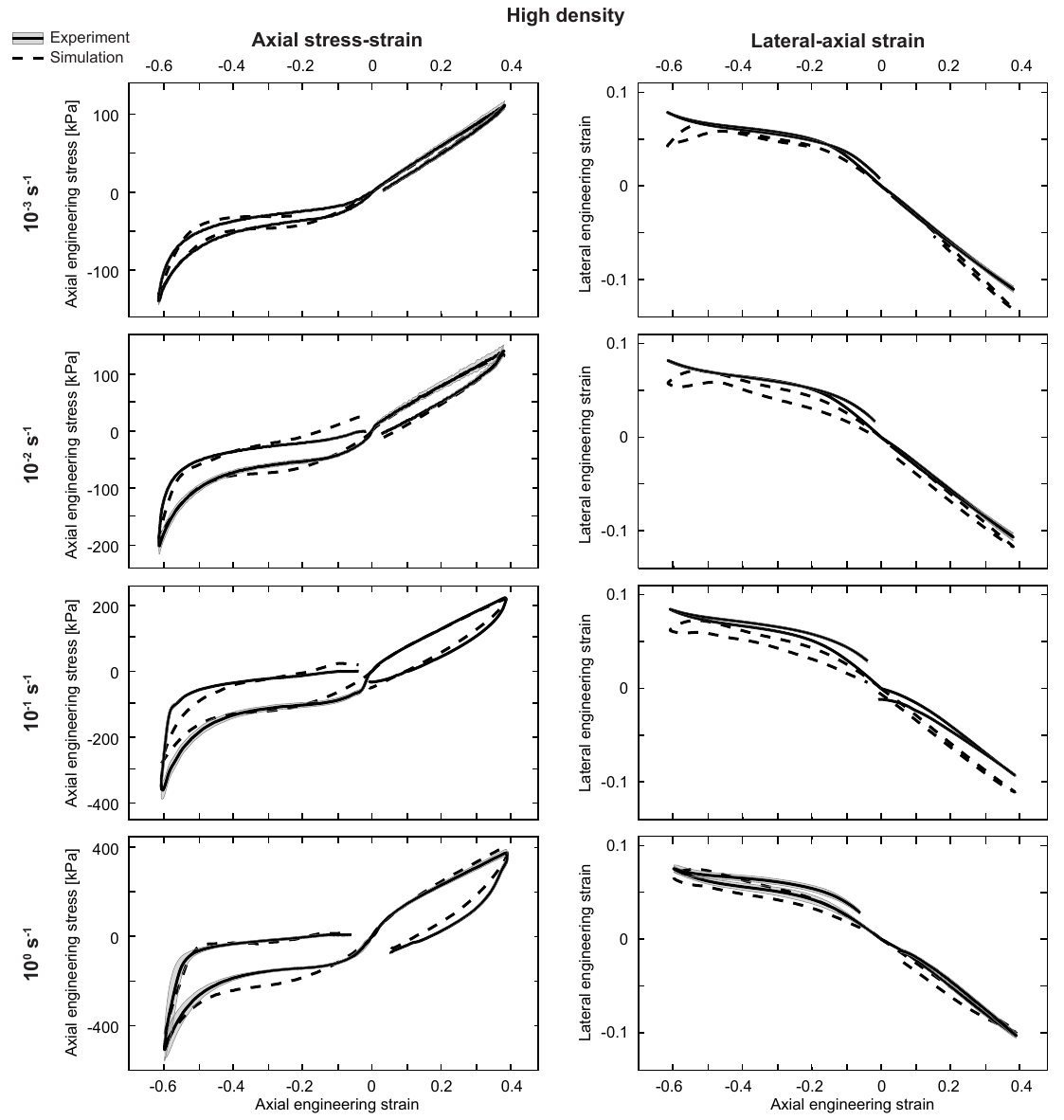}  
	\caption{(\textit{Left}) Axial engineering stress versus axial engineering strain curves and (\textit{Right}) lateral engineering strain versus axial engineering strain curves for the open-cell viscoelastic foam of high density (240\,kg/m$^3$) over the strain-rate range $10^{-3}$\,s$^{-1}$ to $10^0$\,s$^{-1}$. The solid lines with shaded error areas (one standard deviation, three experiments) indicate experimental data, and the dashed lines are constitutive model fits. To best show the characteristics of the curves, the stress axes differ in range for each strain rate.}
	\label{fig:poron155_v_simple}
\end{figure}

The experimentally-measured responses in compression and tension for all four strain rates tested are shown in \autoref{fig:poron95_v_simple}, \autoref{fig:poron125_v_simple}, and \autoref{fig:poron155_v_simple} as solid lines for the low-, moderate-, and high-density open-cell viscoelastic foams, respectively. 
%
%
In each figure, the engineering stress versus axial engineering strain response is shown in the left column, and the corresponding lateral engineering strain versus axial engineering strain response is shown in the right column. The experiment for each case is repeated three times, and the solid lines represent the mean, while the shaded regions surrounding the solid curves represent one standard deviation above and below the mean. The responses for all foam densities and strain rates are highly nonlinear and asymmetric between tension and compression, with four regimes of deformation, as highlighted in the upper left of \autoref{fig:poron95_v_simple}. Moreover, the stress--strain response is rate dependent and displays hysteresis upon reversed loading with the amount of hysteresis and the stress level increasing as the strain rate increases. Importantly, as the strain rate increases, the transition from the linear regime to the plateau regime becomes sharper, especially in the low-density and moderate-density foams. For the moderate-density and high-density foams, the plateau--densification transition occurs at a lower axial strain magnitude for a strain rate of $10^0$\,s$^{-1}$ compared to the lower strain rates. Compared to the closed-cell foam characterized in \citet{LI2022}, all densities of the open-cell foam demonstrate a more significant increase in hysteresis, a larger increase in stress magnitude, and more abrupt transitions between the linear, plateau, and densification regimes as the strain rate increases. These features motivate the changes to the flow rule discussed in \autoref{sec:flowrule} that aid in capturing these effects in the phenomenological modeling framework. 

In the application of the constitutive model to all three densities of the open-cell viscoelastic foam, the one equilibrium branch accounts for the quasistatic response, and $N=12$ non-equilibrium branches are invoked to smoothly capture the responses across strain rates ranging from $10^{-3}$\,s$^{-1}$ to $10^{0}$\,s$^{-1}$ in the experiments. The choice of the number of non-equilibrium branches is an important one, and $N=12$ was empirically determined to balance model performance against the number of adjustable parameters. Importantly, the time constants $t_0^{(\alpha)}$ are not treated as adjustable. Instead, they are \textit{a priori} set for each branch, evenly spanning from $10^{-2}$\,s to $10^{2}$\,s logarithmically, i.e. $t_0^{(\alpha)} = 10^{(4/11)(\alpha-1) - 2}$\,s for $\alpha\in\{1,2,\ldots,12\}$, for the range of strain rates considered in the current experiments. By fixing time constants to be evenly spaced logarithmically over the calibrated range, redundant branches are avoided, and large gaps in the time constants are minimized, which could potentially lead to abrupt jumps in the calibrated viscoelastic stress--strain response as the strain rate is varied. 

Therefore, regarding parameters, we have one equilibrium branch with fourteen material parameters $\{ G_{0}$, $B$, $J_{\rm min}$, $C_1$, $K_1^{0}$, $\Delta_K$, $X_1^\prime$, $X_2^\prime$, $C_0$, $p$, $q$, $C_2$, $C_3$, $r\}$, $N = 12$ non-equilibrium branches each possessing three adjustable parameters $\{G_{\rm neq}^{(\alpha)}, C_{1{\rm neq}}^{(\alpha)}, q^{(\alpha)}_{\rm neq}\}$, and the three branch-independent, viscous-volumetric-strain-dependent viscosity parameters $\{K_{10}^{\rm v}, C_4, C_{\rm e}\}$ for a total of 53 parameters. A systematic, manual method to estimate the parameters for this constitutive model structure exists \citep{LI2022}. However, this manual calibration approach becomes impractical as the number of non-equilibrium branches increases, as the model requires simultaneously tuning a large number of material parameters to achieve consistent fidelity across all four strains rates and across tension and compression for each material of interest. Instead, an extension of the Equilibrium Constrained Genetic Algorithm (ECGA) framework described in \citet{yan2026equilibrium} has been implemented to enable automatic calibration of the constitutive model. The ECGA framework applies a genetic algorithm to minimize an objective function defined through the balances of internal and external forces \citep{flaschel2021unsupervised} with respect to the adjustable material parameters. In this present study, the Averaged Field (AF) functionality of the ECGA framework was employed, which is streamlined for data from experiments involving homogeneous deformation. 
%
%
While the formulation presented in \citet{yan2026equilibrium} is restricted to hyperelasticity, the present implementation incorporates temporal interpolation of the experimental data and explicit time integration of the constitutive model to accommodate history-dependent material behavior. Details of the extended ECGA framework for viscoelasticity will be reported elsewhere. 
The material parameters for all three densities of the open-cell foam were calibrated using the viscoelastic ECGA framework. The hyperelastic, equilibrium-branch material parameters for all three densities are given in \autoref{tab:eq_params}; the branch-dependent, non-equilibrium material parameters are given in \autoref{tab:neq_params}; and the branch-independent, viscous-volumetric-strain-dependent viscosity parameters are given in \autoref{tab:vis_params}. We note that the equilibrium parameters in \autoref{tab:eq_params} are different from the parameters reported in Tab.~1 of \citet{yan2026equilibrium}. This is because, in the present work, the equilibrium and non-equilibrium parameters were simultaneously calibrated to the experimental data across strain rates from $10^{-3}$\,s$^{-1}$ to $10^{0}$\,s$^{-1}$, while in \citet{yan2026equilibrium}, the equilibrium parameters were calibrated to experimental data for a single low strain rate of $2\times10^{-4}$\,s$^{-1}$.

\begin{table}[!t]
\setlength{\tabcolsep}{4.5pt} 
\centering
\begin{tabularx}{155mm}{ccccccccccccccc}
	\toprule
 	\multicolumn{15}{c}{Equilibrium parameters}  \\ 
	\noalign{\smallskip}\hline\noalign{\smallskip}
 \multicolumn{1}{c}{Density}  & \multicolumn{1}{c}{$G_0$ [kPa]} &  \multicolumn{1}{c}{$B$ [kPa]} &  \multicolumn{1}{c}{$J_{\rm min}$} &  \multicolumn{1}{c}{$C_1$}  &  \multicolumn{1}{c}{$K_1^0$} &  \multicolumn{1}{c}{$\Delta_{\rm K}$} &  \multicolumn{1}{c}{$X_1^{\prime}$} &  \multicolumn{1}{c}{$X_2^{\prime}$} &  \multicolumn{1}{c}{$C_0$} &  \multicolumn{1}{c}{$p$} &  \multicolumn{1}{c}{$q$} &  \multicolumn{1}{c}{$r$} &  \multicolumn{1}{c}{$C_2$} &  \multicolumn{1}{c}{$C_3$} \\ 
	\noalign{\smallskip}\hline\noalign{\smallskip}
	Low   & 29.6 & 114.0 & 0.07 & 0.84 & -0.12 & 0.26 & 5.73 & 0.6 & 0.45 & 3.2 & 2.4 & 1.3 & 18.3 & 0.33 \\
	Moderate   & 53.5 & 135.0 & 0.01 & 1.24 & -0.04 & 0.38 & 5.73 & 0.08 & 0.05 & 5.6 & 4.0 & 1.0 & 13.0 & 0.4 \\
      High   & 90.7 & 187.3 & 0.03 & 0.45 & -0.20 & 0.24 & 3.60 & 0.14 & 0.05 & 5.2 & 2.8 & 1.0 & 9.0 & 0.27 \\ 
    \hline
\end{tabularx}
\caption{Equilibrium material parameters for the low-, moderate-, and high-density open-cell viscoelastic foams.} 
\label{tab:eq_params}
\end{table}

\begin{table}[!t]
\setlength{\tabcolsep}{4.5pt} 
\centering
\begin{tabularx}{131mm}{c c | c c c|c c c|c c c}
    \toprule
    \multicolumn{11}{c}{Branch-dependent, non-equilibrium parameters}   \\ 
    \noalign{\smallskip}\hline\noalign{\smallskip} 
    \multicolumn{2}{c}{\,}  & \multicolumn{3}{c}{Low} &  \multicolumn{3}{c}{Moderate} &  \multicolumn{3}{c}{High}\\
	\noalign{\smallskip}\hline\noalign{\smallskip}
    \multicolumn{1}{c}{$\alpha$} & \multicolumn{1}{c}{$t_0^{(\alpha)}$ [s]} & \multicolumn{1}{c}{$G_{\rm neq}^{(\alpha)}$ [kPa]} &  \multicolumn{1}{c}{$C_{\rm 1neq}^{(\alpha)}$} &  \multicolumn{1}{c}{$q_{\rm neq}^{(\alpha)}$} &  \multicolumn{1}{c}{$G_{\rm neq}^{(\alpha)}$ [kPa]} &  \multicolumn{1}{c}{$C_{\rm 1neq}^{(\alpha)}$} &  \multicolumn{1}{c}{$q_{\rm neq}^{(\alpha)}$} & \multicolumn{1}{c}{$G_{\rm neq}^{(\alpha)}$ [kPa]} &  \multicolumn{1}{c}{$C_{\rm 1neq}^{(\alpha)}$} &  \multicolumn{1}{c}{$q_{\rm neq}^{(\alpha)}$}  \\ 
    \noalign{\smallskip}\hline\noalign{\smallskip}

	1  & 100.0 & 2.0 & 1.0 & 5.3     & 6.7 & 4.0 & 2.1           & 8.0 & 1.7 & 3.4  \\
	2  & 43.3 & 4.7 & 5.0 & 2.1     & 4.0 & 5.0 & 2.1          & 22.0 & 4.0 & 4.8  \\
    3  & 18.7 & 0.7 & 4.7 & 4.8     & 2.0 & 2.3 & 2.4          & 22.0 & 0.3 & 6.1  \\ 
    4  & 8.1 & 7.3 & 0.0  &  4.0     & 5.3 & 5.0 & 4.0         & 23.3 & 1.7 & 5.3  \\
	5  & 3.5 & 6.7 & 3.0 &  6.1    & 21.3 & 3.3 & 5.0         & 40.0 & 4.0 & 5.0  \\
    6  & 1.5 & 12.0 & 3.0 & 2.1      & 8.0 & 2.7 & 5.8         & 10.0 & 4.7 & 6.7  \\ 
    7  & 0.66 & 4.7 & 3.3 &  2.9    & 14.0 & 4.3 & 4.5         & 100.0 & 1.3 & 2.9  \\
	8  & 0.29 & 29.3 & 0.7 & 5.6    & 58.7 & 2.0 & 5.3        & 40.0 & 3.0 & 4.5  \\
    9  & 0.12 & 23.3 & 3.3 & 2.9    & 60.0 & 5.8 & 2.0        & 73.0 & 3.0 & 2.1  \\ 
    10 & 0.053 & 53.3 & 3.3 & 5.3    & 26.7 & 0.3 & 2.1        & 220.0 & 0.7 & 2.1  \\
	11 & 0.023 & 40.0 & 5.0 & 3.4      & 167.0 & 0.7 & 3.2          & 100.0 & 2.7 & 4.8  \\
    12 & 0.01 & 106.0 & 1.3 & 5.8     & 240.0 & 2.7 & 5.0          & 267.0 & 3.3 & 4.0  \\ 
    \hline
\end{tabularx}
\caption{Branch-dependent, non-equilibrium material parameters for the low-, moderate-, and high-density open-cell viscoelastic foams.} 
\label{tab:neq_params}
\end{table}

\begin{table}[!t]
\setlength{\tabcolsep}{4.5pt} 
\centering
\begin{tabularx}{82mm}{c c c|c c c|c c c}
	\toprule
 	\multicolumn{9}{c}{Branch-independent, viscosity parameters}   \\ 
  \noalign{\smallskip}\hline\noalign{\smallskip} 
         
  \multicolumn{3}{c}{Low} &  \multicolumn{3}{c}{Moderate} &  \multicolumn{3}{c}{High}\\
  \noalign{\smallskip}\hline\noalign{\smallskip}
	  $C_4$ & $C_{\rm e}$ & $K_{10}^{\rm v}$ & $C_4$ & $C_{\rm e}$ & $K_{10}^{\rm v}$ & $C_4$ & $C_{\rm e}$ & $K_{10}^{\rm v}$ \\
 \noalign{\smallskip}\hline\noalign{\smallskip}
	180.0 & 4.6 & -0.5 & 20.0 & 3.2 & -0.36 & 80.0 & 3.6 & -0.36 \\
    \hline
\end{tabularx}
\caption {Branch-independent, viscous-volumetric-strain-dependent viscosity parameters for the low-, moderate-, and high-density open-cell viscoelastic foams.} 
\label{tab:vis_params}
\end{table}

Using the specialized viscoelastic constitutive model with the calibrated material parameters for the low, moderate, and high-density open-cell viscoelastic foams, the resulting engineering stress--strain response and lateral--axial engineering strain response are plotted as dashed curves alongside the experiments for strains rate of $10^{-3}$\,s$^{-1}$, $10^{-2}$\,s$^{-1}$, $10^{-1}$\,s$^{-1}$, and $10^{0}$\,s$^{-1}$ in \autoref{fig:poron95_v_simple}, \autoref{fig:poron125_v_simple}, and \autoref{fig:poron155_v_simple}. 
The fitted model generally captures the experimental behavior for both the stress--strain and the lateral--axial strain response\,---\,in both the overall shape and the hysteretic behavior upon unloading\,---\,across all densities and strain rates. The largest discrepancies in the axial stress--strain response between the experimental data and the model fit occur in the plateau regime for the moderate- and high-density foams, particularly at the highest strain rate of $10^{0}$\,s$^{-1}$. The worst-case lateral response occurs for the moderate-density foam in compression, also at the $10^{0}$\,s$^{-1}$ strain rate; however, the lateral strains remain small in compression, so the magnitude of this difference is also small. 
%
%
Overall, the quality of comparison is quite reasonable given the level of complexity of the material response across density, strain amplitude, strain rate, and deformation mode (i.e., compression and tension). 

\section{Validation tests}\label{sec:result_validation}

The material model described in \autoref{sec:model} has been numerically implemented in \citet{abaqus} using a user-material (VUMAT) subroutine. See \ref{ap:vumat} for specifics of the VUMAT implementation. To probe the predictive capability of the nonlinear viscoelastic model across inhomogeneous deformation states involving compression, shear, and tension, a series of validation experiments were carried out. The experiments were simulated in \citet{abaqus}, from which predicted variables from the simulations were compared against the corresponding measurements from experiments. Following our prior work \citep{landauer2019quasistatic,LI2022}, three classes of validation tests were undertaken: (1) cyclic conical indentation for all three densities, (2) cyclic simple-shear-like deformation both without and with a fixed amount of pre-compression for the moderate-density foam, and (3) cyclic tension of a specimen with circular holes for the high-density foam. For all validation tests, displacement was the controlled variable and a resultant force was measured. The displacement history for each case was chosen such that the local strain-rate fields achieved in the experiment stayed within the fitted strain-rate range. A spectrum of strain magnitudes, strain rates, and deformation modes were probed across the validation tests to thoroughly evaluate the performance of the model. 

\subsection {Conical indentation} \label{sec:indentation_v_poron}

\begin{figure}[!t]
	\centering\includegraphics[width=0.98\textwidth]{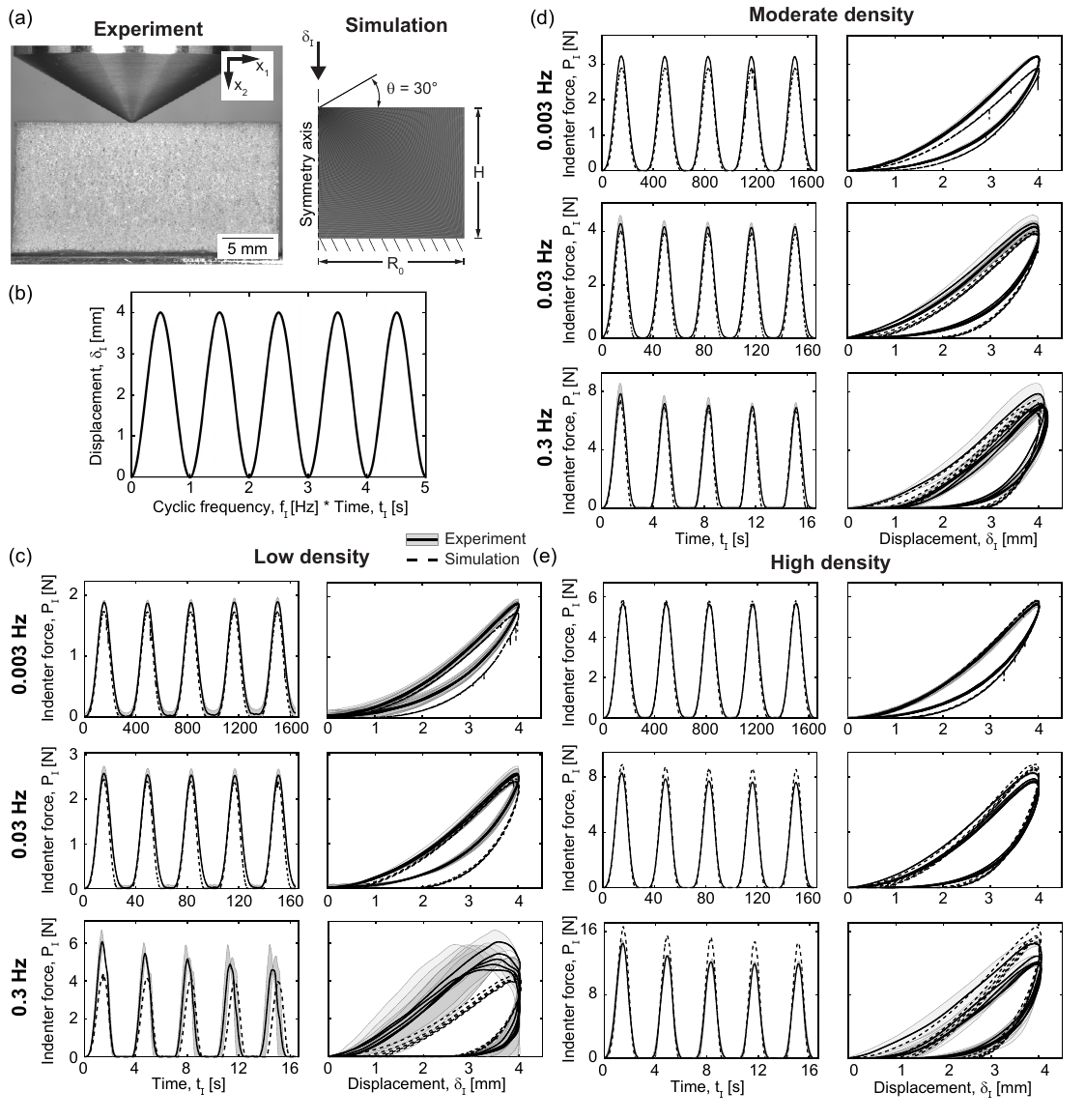}  
	\caption{Conical indentation validation tests for each density of foam. (a) Snapshot of the experimental setup (\textit{Left}) with an undeformed specimen and the indenter tip about to make contact with the specimen, and (\textit{Right}) the corresponding finite-element simulation geometry. (b) Indenter displacement $\delta_{\mathrm I}$ as a function of time normalized by the cyclic frequency of the displacement. (c) Indenter force, $P_{\rm I}$, versus time, $t_{\rm I}$, (\textit{Left}) and corresponding indenter force versus indenter displacement, $\delta_{\rm I}$, (\textit{Right}) for conical indentation of the low-density foam under three different cyclic frequencies of the indenter tip. (\textit{Top}) Low rate, 0.003\,Hz, (\textit{Middle}) moderate rate, 0.03\,Hz, and (\textit{Bottom}) high rate, 0.3\,Hz. Panels (d) and (e) show equivalent results for the moderate- and high-density foams, respectively. Solid lines with shaded error regions are the experimental data, and dashed lines are the simulated predictions obtained from the material model. Note that to best highlight the comparisons for each case, the force axes differ in range for the different testing rates and foam densities.}
	\label{fig:poron_v_indentation}
\end{figure}
 
The experimental setup and geometry used for conical indentation were identical to the configuration described in \citet{landauer2019quasistatic}; however, to probe the viscoelastic response of the materials, the experiments were conducted cyclically over five indentation cycles at one of three cyclic frequencies. To summarize, the conical indentation experiments were conducted on foam specimens for all three densities using a square of nominally 25.4\,mm\,$\times$\,25.4\,mm cross section and a height approximately $H = 12.9$\,mm. Indentation was performed using a 6061 aluminum indenter tip with a base angle of approximately $\theta = 30^{\circ}$. A photograph of the experimental setup prior to deformation is shown in \autoref{fig:poron_v_indentation}(a). Experiments were performed using a sinusoidal indenter displacement profile, as illustrated in \autoref{fig:poron_v_indentation}(b), at three different decadally-spaced cyclic frequencies, $f_{\rm I} = 0.3$\,Hz, $0.03$\,Hz, and $0.003$\,Hz. The maximum indenter displacement during conical indentation was nominally 4\,mm in all cases. 

For the corresponding finite-element simulations, we utilized the same dimensions, boundary conditions, and mesh as described in \citet{landauer2019quasistatic}. Specifically, the simulations were performed in \citet{abaqus} with our custom VUMAT subroutine using an axisymmetric model disk of height $H=12.9$\,mm and radius $R_0=14.4$\,mm with 17\:236 Abaqus-CAX4H elements, as shown in \autoref{fig:poron_v_indentation}(a). The indenter was modeled as a rough, rigid surface; the bottom of the specimen was treated as a fully fixed boundary; and the symmetry axis (left edge in \autoref{fig:poron_v_indentation}(a)) was prescribed to have zero horizontal displacement. All other boundaries of the specimen were traction-free. 

The experimentally-measured indenter force $P_{\rm I}$ versus time $t_{\rm I}$ and indenter force versus displacement $\delta_{\rm I}$ curves for the low-density, moderate-density, and high-density foams are shown as solid lines in \autoref{fig:poron_v_indentation}(c), (d), and (e), respectively. Data for all three cyclic frequencies are shown. The shaded regions surrounding the solid curves represent one standard deviation above and below the mean across three repetitions on the same specimen. The calculated $P_{\rm I}$ versus $t_{\rm I}$ and $P_{\rm I}$ versus $\delta_{\rm I}$ curves are shown as dashed lines for all three frequencies in \autoref{fig:poron_v_indentation}(c), (d), and (e) for each density. In general, the simulation results and the measured experimental data are in agreement throughout all cycles and most rate--density combinations. That said, there is a noticeable discrepancy for the case of the low-density foam at the highest frequency, $f_{\rm I}=0.3$\,Hz. This is likely due to the un-skived nature of these specimens\,---\,i.e., since the top layer of foam is denser and a thin skin is present, the force measured by the indenter will likely be slightly higher than the model prediction due to sink-in and entrapped gas artifacts \citep{HUAJIAN19922471,esposito2020modeling}. This is most significant for the low-density foam, particularly at higher effective strain rates, whereas the moderate- and high-density simulations match well, largely falling within the one standard deviation uncertainty bounds of the experiments.

\subsection {Simple-shear-like deformation without and with pre-compression}\label{sec:shear}

\begin{figure}[!t]
	\centering
	\includegraphics[width=1.0\textwidth]{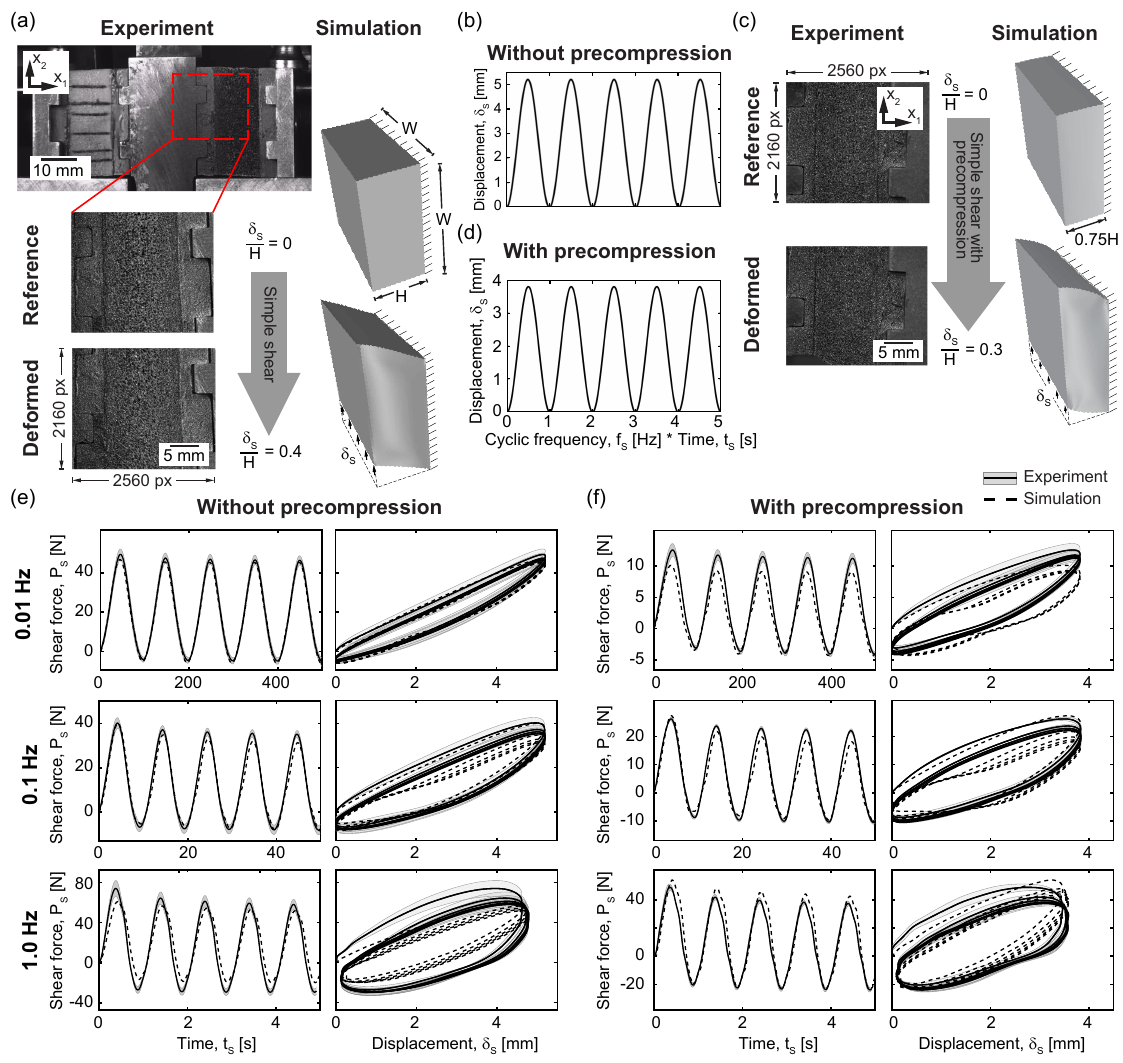}
	\caption{Simple-shear-like validation tests without and with pre-compression for the moderate-density foam. (a) (\textit{Left}) Snapshot of the double-lap shear fixture with two specimens prior to deformation with an inset showing images of the undeformed and deformed specimen surfaces and (\textit{Right}) the corresponding finite-element geometries in the undeformed and deformed states. (b) Displacement history of the center platen as a function of time normalized by the cyclic frequency for the case without pre-compression. (c) (\textit{Left}) Images of a specimen in the pre-compressed and the deformed, sheared states and (\textit{Right}) renders of the corresponding finite-element geometries. (d) Displacement history of the center platen as a function of time normalized by the cyclic frequency for the case with pre-compression. (e) (\textit{Left}) Shear force, $P_{\rm S}$, versus time, $t_{\rm S}$, and (\textit{Right}) $P_{\rm S}$ versus shear displacement, $\delta_{\rm S}$, for three cyclic frequencies without pre-compression. (f) (\textit{Left}) $P_{\rm S}$ versus $t_{\rm S}$ and (\textit{Right}) $P_{\rm S}$ versus $\delta_{\rm S}$ for three cyclic frequencies with pre-compression. Solid lines with shaded error regions represent the experimental data, and dashed lines are the simulated predictions obtained from the material model. Note that to best highlight the comparisons for each case, the force axes differ in range for the different testing rates and pre-compression states.}
	\label{fig:shear_v_poron}
\end{figure}

As another validation test, predictions of the viscoelastic model were also compared against experimental measurements in two shear-dominated deformations for the moderate-density foam: double-lap shear and double-lap shear with pre-compression. The experiments and simulations use the same setups as described in \citet{landauer2019quasistatic} with cyclic simple-shear-like experiments performed on foam cuboids with a square cross-section of nominal side-length $W = 30$\,mm and height $H=12.9$\,mm. A snapshot of the double-lap shear fixture with two undeformed specimens is shown in \autoref{fig:shear_v_poron}(a). The shear fixture was installed in the load frame used for the low strain-rate characterization experiments to apply the shear displacement. For shear with pre-compression, the specimens were first compressed, so that the deformed distance between the rectangular faces was $0.75H$. Then, the specimens were allowed to relax for one or more hours between pre-compression and subsequent application of the shear displacement history. The shear displacement $\delta_{\rm S}$ was applied with a sinusoidal history at three different cyclic frequencies, $f_{\rm S} = 1.0$\,Hz, $0.1$\,Hz, and $0.01$\,Hz. The input displacement profiles for shear without and with pre-compression are shown in \autoref{fig:shear_v_poron}(b) and (d), respectively. The maximum shear displacement was $\delta_{\rm S}/H = 0.4$ for the case without pre-compression and $\delta_{\rm S}/H = 0.3$ for the case with pre-compression. Photographs of the reference and sheared specimens without and with pre-compression are shown in \autoref{fig:shear_v_poron}(a) and (c), respectively. 

Three-dimensional finite-element simulations were carried out on a single cuboidal specimen consisting of 97\:000 Abaqus-C3D8H elements. Displacement boundary conditions on the fixed-side backer were set to zero for all components, and for the moved-side backer, the vertical displacement history $\delta_{\rm S}$ was prescribed, while the remaining components were set to zero. For the pre-compression condition, the moving-side backer was first prescribed a quasistatic displacement of $u_1 = -0.25H$ prior to imposing the vertical displacement history $\delta_{\rm S}$. All lateral faces of the specimen were traction-free for cases both without and with pre-compression. \autoref{fig:shear_v_poron}(a) shows the undeformed mesh and the deformed mesh after shear, and \autoref{fig:shear_v_poron}(c) shows the deformed meshes after pre-compression and subsequent shear. 

The experimentally measured shear force $P_{\rm S}$ for one specimen versus time $t_{\rm S}$ histories and the shear force versus shear displacement $\delta_{\rm S}$ relations for all three frequencies without and with pre-compression are shown as solid lines in \autoref{fig:shear_v_poron}(e) and (f), respectively. The shaded regions surrounding the solid curves represent one standard deviation above and below the mean across three repetitions on the same specimen. 
%
%
The simulation results for shear force versus time and shear force versus shear displacement for each frequency are included as dashed lines in \autoref{fig:shear_v_poron}(e) for shear without pre-compression and in \autoref{fig:shear_v_poron}(f) for shear with pre-compression. The model captures the main features of the response both without and with pre-compression across loading frequencies typically to within the bounds of the experimental uncertainty. In particular, the model predicts a stiffer, more hysteretic force--displacement relation as the loading frequency increases, which is consistent with experimental data. Moreover, the model captures the effect of pre-compression, in which the force–displacement relation becomes more compliant with pre-compression, which is captured by the model through the coupling between distortional and volumetric deformation in the free-energy density function \autoref{eq:free_energy}. This effect is observed for all three loading frequencies but is strongest for the lowest frequency. The model can capture this rate-dependence because the non-equilibrium free-energy density function for branch $\alpha$, $ \tilde\psi_{\rm neq}^{(\alpha)}(K_1^{{\rm e}(\alpha)},K_2^{{\rm e}(\alpha)},K_3^{{\rm e}(\alpha)})$, depends on the respective non-equilibrium elastic volumetric strain $K_1^{{\rm e}(\alpha)}$ and not the total volumetric strain $K_1$.  
Since complete viscoelastic relaxation is allowed between pre-compression and shearing, $K_1^{{\rm e}(\alpha)} = 0$ for all non-equilibrium branches at the beginning of shearing, and $K_1^{{\rm e}(\alpha)}$ remains near zero in all non-equilibrium branches during shear deformation. Therefore, the non-equilibrium branches do not contribute to the increased compliance with pre-compression, leaving the equilibrium branch to account for distortional/volumetric coupling, which has a stronger relative contribution to the response at lower loading frequencies. 
As the loading frequency increases, the viscous behavior becomes more dominant; the non-equilibrium branches contribute more to the total resultant shear force; and the relative contribution of the distortional/volumetric coupling stemming from the equilibrium branch becomes less strong. 
The model also can also generally capture the higher-force response of the foam during the first cycle of pre-compressed loading and subsequent cyclic softening to a stabilized hysteresis loop, although it does not capture this effect for the highest frequency case without pre-compression. 

\subsection{Tension of a specimen with circular holes}\label{sec:inhomo_tension_vp}

\begin{figure}[!t]
	\centering
	\includegraphics[width=1.0\textwidth]{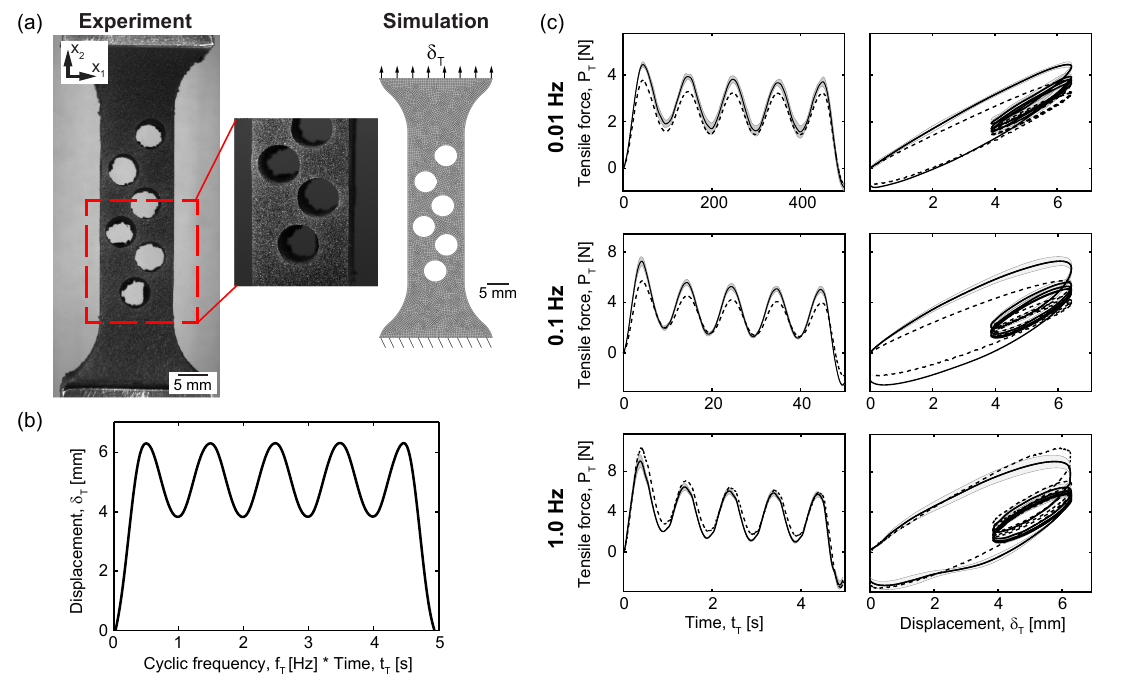}
	\caption{A tension-dominated validation test for the high-density foam. (a) Snapshot of the undeformed specimen with a closeup of part of the gauge section and the corresponding finite-element configuration. (b) Displacement history of the top grip as a function of time normalized by the cyclic frequency. (c) (\textit{Left}) Top-grip force, $P_{\rm T}$, versus time, $t_{\rm T}$, and (\textit{Right}) $P_{\rm T}$ versus top-grip displacement, $\delta_{\rm T}$, for cyclic frequencies of 0.01\,Hz, 0.1\,Hz and 1.0\,Hz. Solid lines with shaded error regions represent the experimental data, and dashed lines are the simulated predictions obtained from the material model. Note that to best highlight the comparisons for each case, the force axes differ in range for the different testing rates.}
	\label{fig:inhomo_tension_v_poron}
\end{figure}

As a final validation test, we compare model predictions to experimental results for a tensile specimen with circular holes cut in the gauge section, focusing on the high-density foam. 
The experimental setup and specimen geometry used for this tension-dominated validation are, again, as described in \citet{landauer2019quasistatic}. A photograph of the specimen installed in the load frame is shown in \autoref{fig:inhomo_tension_v_poron}(a) with a closeup of part of the gauge section with circular holes. The specimen was subjected to five cyclic loading histories under tension, i.e., positive force-ratio cyclic loading. The displacement history $\delta_{\rm T}$ is shown in \autoref{fig:inhomo_tension_v_poron}(b) as a function of normalized time. In this configuration, the bottom grip was fixed and the top grip moved at three different cyclic frequencies: 0.01\,Hz, 0.1\,Hz and 1.0\,Hz. The 0.01\,Hz and 0.1\,Hz experiments were carried out using the load frame used for the low-strain-rate experiments, and the 1.0\,Hz experiment was carried out using the load frame used for the high-strain-rate experiments. The corresponding three-dimensional finite-element model of the specimen consisted of 185\:216 Abaqus-C3D8H elements, as shown in \autoref{fig:inhomo_tension_v_poron}(a). The bottom of the mesh was fully constrained; the vertical displacement history $\delta_{\rm T}$ was imposed on the top face with the other displacement components set to zero; and all other faces were traction-free.  

The measured tension force $P_{\rm T}$ versus time $t_{\rm T}$ and $P_{\rm T}$ as a function of top-grip displacement $\delta_{\rm T}$ are shown in \autoref{fig:inhomo_tension_v_poron}(c) as solid lines with shaded error regions. As before, the shaded area denotes the range of one standard deviation above and below the mean experimental measurement.
In the experiments, the peak force in the first cycle is substantially higher than in subsequent peaks for all rates, followed by cyclic softening to a stabilized hysteresis loop. This effect is well-captured by the model in the highest-rate case but not as well for the lower rates. The stress level and amount of hysteresis of the stabilized loop increases with loading frequency, which is captured well by the model. Due to the viscoelastic nature of the foams, in the final unloading segment of the displacement history, the material goes into compression, due to relaxation during cyclic loading at a positive force ratio, which is also captured well by the model, in particular for the $0.01$\,Hz and $1.0$\,Hz cases. Overall, the performance of the constitutive model is often within the experimental uncertainty, in particular for the $1.0$\,Hz case, although with under-prediction of the maximum forces for the $0.1$\,Hz and $0.01$\,Hz cases. 

\section{Strain localization analysis}\label{sec:loc}

The axial and lateral engineering strains in the calibration experiments were calculated by a global plane-fitting technique to minimize noise \citep{landauer2018qdic}. This approach is based on the assumption of homogeneous deformation in the specimen, which was verified to be a reasonable idealization for a low strain rate of $2\times 10^{-4}$\,s$^{-1}$ in \citet{landauer2019quasistatic}. However, as the strain rate increases this assumption might be violated as a shift in deformation habits from homogeneous compression to the emergence of compaction bands occurs, similar to the length-scale-related discussion of \citet{koohbor_multiscale_2021}. This change in behavior is often associated with a transition from a monotonic to a non-monotonic engineering stress--strain response. 
%
%
These features are qualitatively associated with collective buckling during strain localization or other meso-scale geometric nonlinearities. Since localization is defined through the spatial distribution of strain, a direct measure of localization intensity is obtainable from the full-field strain measurements. Observations of histograms of the strain field provide a semi-quantitative method, as in \citet{wang2002} and \citet{landauer2019quasistatic}. Here, we introduce and use a new quantitative method to measure localization intensity (LI) for a sample of representative experiments at and above the rates used for model calibration.

The localization quantification algorithm directly uses the spatial frequency content of the surface strain field to infer the LI. A convolutional spatial-differentiation filtering kernel is used to compute strain fields from the DIC displacement fields with kernel coefficients optimized to maximize signal and minimize noise, as described in \citet{BarKochba2015}, \citet{landauer2019quasistatic}, and \citet{farid2004}. For a specimen that does not exhibit localization, the mean strain calculated by spatially averaging the strain field generally agrees to within one standard deviation with the strain inferred from a global plane fit. A 9-tap filter was used in this analysis, which provides a spatial resolution of roughly \SI{250}{\micro\metre}, which is similar to the length-scale of the cellular structure of the foams described in \autoref{sec:exp}. Once the strain fields have been computed, an isotropic wavelet of the Marr type, conceptually similar to a Laplacian-of-Gradients-type blob detection, with $\mathrm{L}^2$-normalization is applied at several length scales to the component of the strain field aligned with the compression axis, here the $\mathbf{e}_1$-direction and $E_{11}$ Green-Lagrange strain component. Thresholding is applied to the wavelet response to identify spatial regions where the spatial frequency content suggests the presence of strain localization. Denoting the mean strain in the localization and non-localization regions of the strain field as $\bar{E}_{11_{\rm loc}}$ and $\bar{E}_{11_{\rm non\,loc}}$, respectively, and their standard deviations as ${\rm SD}_{\rm loc}$ and ${\rm SD}_{\rm non\,loc}$, the LI and its uncertainty are then defined as the absolute difference between the mean strains in the localization and non-localization regions and the $\mathrm{L}^2$-norm of their standard deviations: 
\begin{equation}
    {\rm LI} = \left||\bar{E}_{11_{\rm loc}}| - |\bar{E}_{11_{\rm non\,loc}}|\right| \pm \sqrt{{\rm SD}_{\rm loc}^2 + {\rm SD}_{\rm non\,loc}^2}.
\end{equation} 
%
%
The LI quantifies the level of excess strain above the mean global strain in the regions where the spatial frequency content of the strain map suggests localization may be occurring. The complete code to perform this analysis, qDIC-based Strain Localization Analysis with Wavelets (Q-SLAW), is available as Matlab scripts in \citet{localization_code}.


\begin{figure}[!t]
	\centering{\includegraphics[width=1\textwidth]{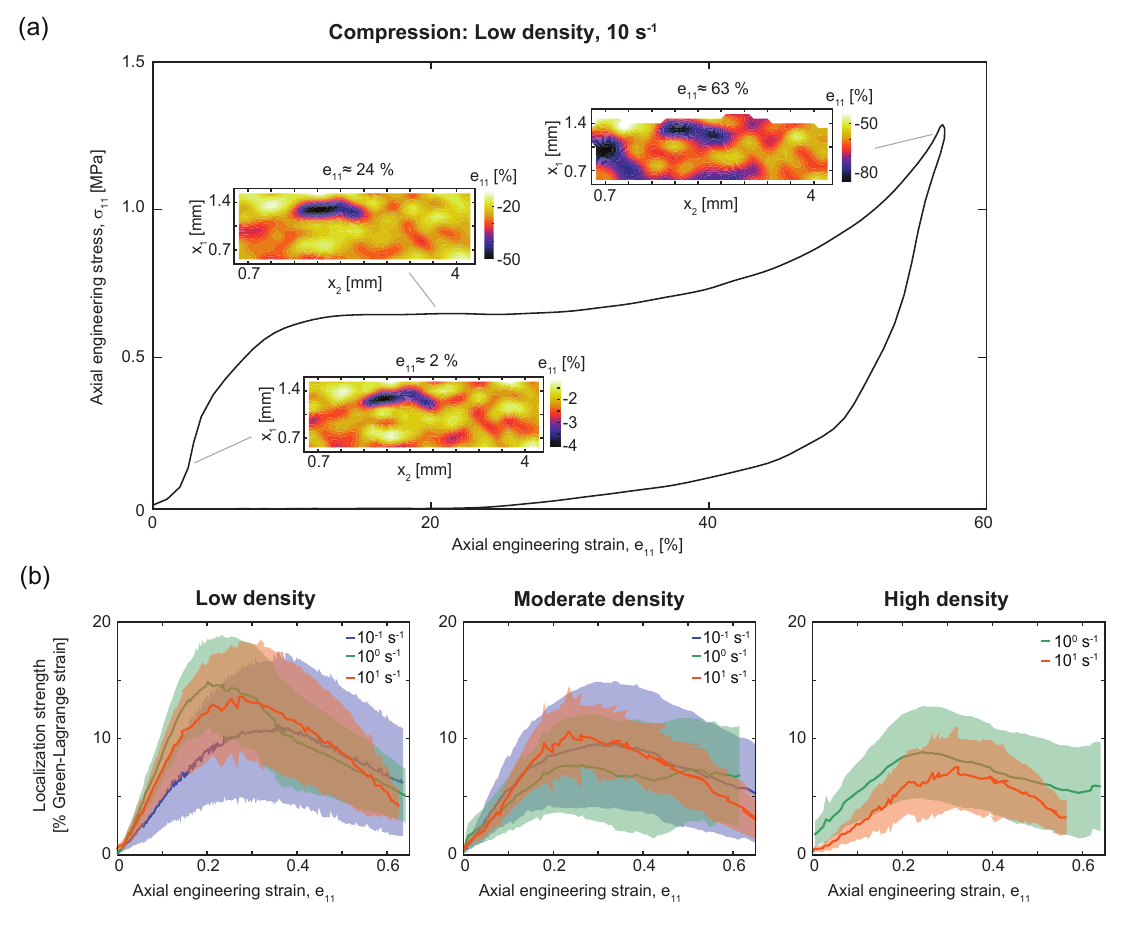}}
	\caption{A representative compression experiment on the low-density foam at a strain rate of $10^1$\,s$^{-1}$, demonstrating some signs of localization at the impact-like rate. (a) Engineering stress--strain plot with representative maps of the surface engineering strain fields for the three deformation regimes in compression. Instrument uncertainty associated with stress measurement is $\pm$8.9\,kPa. (b) The Localization Intensity (LI) metric with propagated uncertainty (shaded regions) as a function of the imposed axial engineering strain at assorted strain rates for the low-, moderate-, and high-density foams. Note that engineering strain is used for visualization, whereas Green-Lagrange strain is used in the analysis.}
	\label{fig:c095_localization}
\end{figure}

The strain maps provide insights into the deformation behavior at different stages of compression, as illustrated in \autoref{fig:c095_localization}(a) for the low-density foam at a strain rate of $10^1$\,s$^{-1}$. While the viscoelastic constitutive model was not applied to this case, it is anticipated to be the most prone to strain localization among the density and strain-rate combinations considered in this work. Initially, in the linear elastic regime, a small hotspot in the strain field begins to form. As compression progresses beyond the transition into the plateau regime, the strain hotspot intensifies, and a secondary, less intense spot begins to form on the middle left. Finally, in the densification regime, the secondary hotspot has grown to match the initial hotspot more closely in size and intensity. To quantify these observations, the LI as a function of axial engineering strain, calculated using Q-SLAW, is shown in \autoref{fig:c095_localization}(b) for all three foam densities at several of the higher strain rates considered here. The LI increases until approximately the onset of densification, when the LI begins to reduce. In agreement with empirical observations, the LI is lower for higher relative density foams, and typically higher for higher strain rates. However, for these foam densities at these strain rates, the differences are largely within the uncertainty bounds, with the exception of the high-density foam. Since strain inhomogeneity remains relatively limited in the calibration experiments, i.e., ${\rm LI}\lesssim 0.15$, the idealization of homogeneous deformation remains valid.


\section{Limitations}\label{sec:limitations}
For model development and the corresponding validation experiments, we considered the foams to be homogeneous and isotropic materials, neglecting the influence of the micro- and meso-scale structure on the continuum-scale material mechanics. Some indication of strain localization due to collective buckling was measured via the localization analysis of \autoref{sec:loc}. Thus, including the micro- and meso-scale mechanisms of ligament buckling or node rotation quantified by meso-scale measurement tools such as those of \citet{rouwane_architecture-driven_2022} into the model\,---\,e.g., using a nonlocal continuum formulation such as the one described by \citet{JOSHI2026118719}\,---\,could help predict deviations due to localization. This would improve the predictive capability and broaden the range of foams and strain rates that can be successfully addressed by the model. Additionally, although the material in this work was isotropic, many blown foams exhibit anisotropy, e.g., along a rise-direction, and an extension of the modeling framework would be needed to capture anisotropy. Also, the upper limit of the calibrated strain-rate range in the present work was $10^{0}$\,s$^{-1}$, which limits the range of impact scenarios that may be successfully modeled with the calibrated model. 

Finally, a comment on the large number of material parameters invoked in the current model is warranted, which at first glance might seem excessive. However, it is important to recognize that the physical behavior of this class of materials is highly nonlinear with significant tension--compression asymmetry, loading--unloading hysteresis, and strong deviatoric--dilatational coupling, while also featuring substantial material compressibility. Invoking a greater number of non-equilibrium branches ($N=12$) with relaxation times that span the timescales of the experiments logarithmically leads to higher fidelity in simulations of scenarios involving strain rates different from the experiments\,---\,i.e., the interpolative capability of the model is improved.

Alternatively, machine-learning-based approaches have demonstrated capability for constitutive modeling of elastomeric foams \citep[e.g.,][]{liang2008neural,frankel2022machine}. For finite-deformation viscoelasticity, related approaches have been developed to represent history-dependent constitutive behaviors directly from data or to replace portions of conventional constitutive models with learned functions while satisfying physical and thermodynamic restrictions \citep[e.g.,][]{tacc2023data,abdolazizi2024viscoelastic,kalina2026physics}. Such approaches can potentially provide greater flexibility and reduce the need to prescribe phenomenological functional forms. However, the resulting predictive capability still depends on the scope of the material response represented by the available experimental data and on the constitutive restrictions imposed on the learned response, while additional challenges arise in ensuring physical admissibility, numerical robustness, and implementation. 

The present work pursued a classical constitutive formulation, and while the model involves a considerable number of mechanisms and associated material parameters, the choice in the type and number of material parameters was determined by balancing maintaining the smallest number and most interpretable set of parameters with good predictive ability. Current and future efforts will continuously evaluate the number of required degrees of freedom in the constitutive model to remain predictive across a wide gamut of homogeneous, inhomogeneous, and multi-axial loading experiments, as has been demonstrated here. In an effort to enable access to the model across a wide user base, the numerical implementation in an \citet{abaqus} VUMAT subroutine has been made available for download. 

\section{Concluding remarks}\label{conclusion}

\subsection{Summary}
This study introduced a phenomenological constitutive model for the large-deformation, rate-sensitive response of nominally isotropic and homogeneous open-cell viscoelastic foams. To characterize the rate-dependent mechanical behavior of these elastomeric foams, a series of homogeneous simple compression and tension experiments were performed for three densities over decadal strain rates over a range from $10^{-3}$\,s$^{-1}$ to $10^{0}$\,s$^{-1}$, inclusive. A new strain localization analysis algorithm was developed and demonstrated that strain inhomogeneity remained relatively minor. The engineering stress--strain and lateral--axial strain responses of the materials measured from the nominally homogeneous experiments served as inputs for informing and refining the phenomenological, large-deformation, isotropic, viscoelastic constitutive model, which was then implemented as an \citet{abaqus} user-material subroutine and is available in a code repository. After calibration, the model successfully captured the engineering stress versus axial strain and lateral strain versus axial strain responses for each of the three densities foams over the range of strain rates considered experimentally. Specifically, the calibrated model captures a range of significant constitutive features exhibited by elastomeric foams, including the hysteresis during loading and unloading and in cyclic loading, the asymmetry observed between the compression and tension responses, the linear--plateau--densification features in compression, the volumetric--deviatoric coupling, and the influence of strain rate on the stress response. We evaluated the constitutive model via a series of validation experiments conducted under three distinct types of inhomogeneous deformation: conical indentation, simple-shear-like deformation without and with pre-compression, and tension tests on a specimen with circular holes. These validation experiments encompassed a range of local material strain magnitudes, strain rates, and deformation modes, which assessed the model's predictive capacity across the calibrated range. The comparison between the model predictions and experimental data indicated substantial agreement, demonstrating the efficacy of the modeling approach in capturing the nonlinear viscoelastic behavior of foams across diverse loading scenarios and deformation rates. 

\subsection{Future work}
Extending the application of the model to higher strain rates via, for example, Kolsky bar \citep{koumlis2019strain,bhagavathula_density_2022} or drop tower \citep{landauer_stress_2020,bailly_strain_2020,morrison2024temperature} testing to acquire data, is an obvious next step. Another important aspect for future investigation is the inclusion of temperature effects on the mechanical behavior of viscoelastic foams \citep{morrison2024temperature,tao2024experimental}. We briefly established the temperature dependence of the small-strain linear viscoelastic response of the foams considered here; however, the model is isothermal and only calibrated at room temperature. In real-world applications, impact protection products undergo tens of degrees Celsius of temperature change during use, and while these temperature changes are modest, they occur near the glass transition temperature, at which the material exhibits its greatest temperature sensitivity. Addressing this deficiency in the present modeling framework would require conducting calibration experiments across a range of temperatures to investigate the temperature dependence of the response functions in the constitutive model.

\section*{Author contributions statement}

JT and AL conceived, conducted, and analyzed calibration and validation experiments. AL conducted DMA experiments and developed the localization analysis with input from CF, JT, and ZY. ZY and XL conceived and conducted modeling, model calibration, and validation simulations with input from DH. JT, AL, ZY and DH wrote the manuscript and prepared the figures. CF and DH oversaw the experimental and modeling programs, respectively, and edited the manuscript. All authors reviewed the final draft. 

\section*{Declaration of competing interest} %
The authors declare that they have no known competing financial interests or personal relationships that could have appeared to influence the work reported in this paper.

\section*{Acknowledgements}
This work was partially supported by funds from the US Army Research Office under grant W911NF-16-1-0084 and the US Office of Naval Research under PANTHER award number N00014-21-1-2044 through Dr. Timothy Bentley. AKL acknowledges early support from an NSF Graduate Research Fellowship (DGE 1058262), and a portion of the work was conducted while he held NRC Research Associateship awards at the National Institute of Standards and Technology. Partial funding was provided by NIST. We engaged in fruitful discussions with Dr. Aaron M Forster and Dr. Stylianos Koumlis. Portions of \autoref{fig:strain_strainrate_plots}, \autoref{fig:poron_v_indentation}, \autoref{fig:shear_v_poron}, and \autoref{fig:inhomo_tension_v_poron} were adapted with permission from \citet{landauer2019dissertation}.

\section*{Data Availability}\label{sec:data_availablity}

The complete data (model calibration and validation, figure data) will be made available from the NIST Public Data Repository upon journal aceptance or from the authors upon reasonable request. The \cite{abaqus} VUMAT subroutine and input files used in \autoref{sec:result_validation} are available on GitHub at  https://github.com/HenannResearchGroup, and the Q-SLAW code used to perform the localization detection and analysis is available via the NIST Public Data Repository or GitHub: https://doi.org/10.18434/mds2-3141 or https://github.com/usnistgov/Q-SLAW.

%


\appendix

\renewcommand\thetable{A.\arabic{table}} 
\section{Representative DIC information}\label{app_DIC}
\setcounter{figure}{0}
\setcounter{table}{0}

For digital image correlation (DIC) analysis, about 200 images were taken in sequence during the load--unload procedure in each experiment and processed incrementally before cumulative displacements were computed in a finite-deformation framework. DIC specification details for typical compression/tension experiments are given in \autoref{tab:dic_table}. The noise floor was estimated from the standard deviation of sequential, nominally zero-motion images, and the spatial resolution was taken to be half the minimum subset size, which is broadly consistent with experimental observations of the spatial frequency reconstruction capability of the algorithm implementation \citep{landauer2018qdic}.

\begin{table}[!t]
\centering
	 \begin{tabularx}{0.86\textwidth}{c|cccc}
	 \toprule
Specification & Low-rate & Low-rate & Moderate-rate & Moderate-rate \\
 & compression & tension & compression & tension\\
 \hline
Technique& \multicolumn{4}{c}{Single Camera q-Factor Based DIC (2D-DIC) \citep{landauer2018qdic}}\\
Pre-filtering& \multicolumn{4}{c}{Gaussian, [3\,px, 0.2]}\\
Min. Subset Size [px] & \multicolumn{4}{c}{16}\\
Min. Step Size [px] & \multicolumn{4}{c}{8}\\
Reference image mode & \multicolumn{4}{c}{Incremental correlation}\\
Correlation criterion &\multicolumn{4}{c}{FFT-based ZNCC}\\
Interpolation & \multicolumn{4}{c}{Bicubic splines}\\
Camera & \multicolumn{4}{c}{Photron FASTCAM Mini AX-200} \\
Lens & \multicolumn{4}{c}{ZEISS Milvus 2/100 m with Bower SX7DGN teleconverter} \\
Camera noise (est.) & 0.23\,\% & 0.30\,\% & 0.37\,\% & 0.34\,\%\\
Image size [px$^2$] & 1100\:$\times$\:450 & 650\:$\times$\:1000 & 550\:$\times$\:270 & 330\:$\times$\:670\\
Measurement points & 7700 & 10100 & 2300 & 3400\\
Pixel-to-$\mu$m conversion&1\,px\,=\,\SI{9}{\micro\metre} & 1\,px\,=\,\SI{10}{\micro\metre} & 1\,px\,=\,\SI{20}{\micro\metre} & 1\,px\,=\,\SI{25}{\micro\metre}\\
Resolution: spatial (est.) & 8\,px\,=\,\SI{72}{\micro\metre} & 8\,px\,=\,\SI{80}{\micro\metre} & 8\,px\,=\,\SI{160}{\micro\metre} & 8\,px\,=\,\SI{200}{\micro\metre}\\
Resolution: noise [px] & 0.022 & 0.47 & 0.017 & 3.0\\
 \hline
\end{tabularx}
\caption{Typical DIC specification details for simple compression/tension experiments using the low-strain-rate and moderate-strain-rate setups.} 
\label{tab:dic_table}
\end{table}

\renewcommand\thetable{B.\arabic{table}} 
\section{Material parameters ranges}\label{app_GA}
\setcounter{figure}{0}
\setcounter{table}{0}
The model was calibrated for each of the three foam densities using a viscoelastic generalization of the Equilibrium Constrained Genetic Algorithm (ECGA) framework of \citet{yan2026equilibrium}. In this calibration process, the genetic algorithm  requires the specification of a range for each material parameter being estimated. The ranges employed for the non-dimensional, equilibrium-branch parameters\,---\,i.e., the equilibrium parameters other than $G_0$ and $B$\,---\,and the ranges for the non-dimensional, non-equilibrium-branch parameters\,---\,i.e., other than $G_{\rm neq}^{(\alpha)}$\,---\,are shown in \autoref{tab:prescribed}. These ranges are fixed across all foam densities and, in our experience, are wide enough that they can accommodate the behavior of other foam materials. The ranges for $G_0$ and $B$ were calculated from plus and minus 50\% of the $G_0$ and $B$ estimated from the linearized Young's modulus and Poisson's ratio extracted from the engineering stress--strain and lateral--axial strain curves in the linear regime. The lower bounds for the non-equilibrium shear moduli $G_{\rm neq}^{(\alpha)}$ were always chosen to be zero, while the upper bounds scaled from $1/4$ of the upper bound of $G_0$ to four times the upper bound of $G_0$ as $\alpha$ increased from 1 to 12. However, there is not a strict guide for prescribing the ranges of $G_{\rm neq}^{(\alpha)}$. If the number of non-equilibrium branches were chosen to be fewer, the upper bound for each $G_{\rm neq}^{(\alpha)}$ should be cautiously increased and vice versa. The genetic algorithm in the ECGA framework employs binary encoding of the material parameters, and four-bit precision was used throughout this work\,---\,i.e., the precision for each material parameter was its prescribed range divided by $2^4=16$. 

\begin{table}[!t]
\setlength{\tabcolsep}{4.pt} 
\centering
	\begin{tabularx}{161mm}{cccccccccccccccccc}
	\toprule
 \multicolumn{1}{c}{Parameter range} &  \multicolumn{1}{c}{$J_{\rm min}$} &  \multicolumn{1}{c}{$C_1$}  &  \multicolumn{1}{c}{$K_1^0$} &  \multicolumn{1}{c}{$\Delta_{\rm K}$} &  \multicolumn{1}{c}{$X_1^{\prime}$} &  \multicolumn{1}{c}{$X_2^{\prime}$} &  \multicolumn{1}{c}{$C_0$} &  \multicolumn{1}{c}{$p$} &  \multicolumn{1}{c}{$q$} &  \multicolumn{1}{c}{$r$} &  \multicolumn{1}{c}{$C_2$} &  \multicolumn{1}{c}{$C_3$} &  \multicolumn{1}{c}{$C_4$} &  \multicolumn{1}{c}{$C_{\rm e}$} &  \multicolumn{1}{c}{$K_{10}^{\rm v}$} & \multicolumn{1}{c}{$C_{\rm 1neq}^{(\alpha)}$} &  \multicolumn{1}{c}{$q_{\rm neq}^{(\alpha)}$} \\ 

	\noalign{\smallskip}\hline\noalign{\smallskip}
      min  & 0.1 & 0.5 & 0.0 & 0.05 & 2.0 & 0.01 & 0.01  & 2 & 2 & 1 & 5 & 0.001 & 0 & 2 & -0.01 & 0 & 2.1\\
	  max & 0.3 & 6.0 & -0.5 & 0.4 & 10.0 & 1.0 & 6.0 & 8 & 8 & 6 & 25  & 0.5 & 300 & 5 & -0.5 & 5 & 6.1 \\
	\hline
\end{tabularx}
\caption{Prescribed ranges for the non-dimensional material parameters used in the viscoelastic generalization of the ECGA framework \citep{yan2026equilibrium}.} 
\label{tab:prescribed}
\end{table}

\section{Abaqus VUMAT implementation}\label{ap:vumat}
The constitutive model of \autoref{sec:model} was implemented in an \citet{abaqus} user-material subroutine (VUMAT). The state variables of the VUMAT are the viscous deformation gradient ${\bf F}^{{\rm v}(\alpha)}$ and the Mandel stress ${\bf M}^{{\rm e}(\alpha)}$ for each non-equilibrium mechanism $\alpha$. Consider a time $t=t_n$, at which ${\bf F}^{{\rm v}(\alpha)}_n$ and ${\bf M}^{{\rm e}(\alpha)}_n$ for each $\alpha$ are known. Then, given a time increment $\Delta t$ as well as ${\bf F}_{n+1}$ at time $t_{n+1}=t_n + \Delta t$, we are to calculate the Cauchy stress ${\bf T}_{n+1}$ and the internal variables for each $\alpha$, ${\bf M}^{{\rm e}(\alpha)}_{n+1}$ and ${\bf F}^{{\rm v}(\alpha)}_{n+1}$, at time $t_{n+1}$. For the equilibrium branch, the spatial logarithmic strain ${\bf E}_{n+1}$ and its invariants $\{K_{1,n+1}, K_{2,n+1}, K_{3,n+1}\}$ are computed from the deformation gradient ${\bf F}_{n+1}$ to calculate the equilibrium contribution to the Cauchy stress ${\bf T}_{{\rm eq},n+1}$ at time $t_{n+1}$, using \autoref{eq:cauchy2}, \autoref{eq:coupled_fit}, \autoref{eq:L_1}, and \autoref{eq:f_vr}. The non-equilibrium branches are uncoupled and are each updated using an explicit time-integration procedure, as follows.
\begin{enumerate}
\item[\bf Step 1.] Calculate the viscous stretching for mechanism $\alpha$ at time $t_n$ using \autoref{eq:Dv}:
\begin{equation}
\bfD_n^{{\rm v}(\alpha)} = \dfrac{\dev(\bfM_n^{{\rm e}(\alpha)})}{2\eta^{(\alpha)}(K_{1,n}^{{\rm v}(\alpha)})} + \dfrac{\Tr(\bfM_n^{{\rm e}(\alpha)})}{9\kappa^{(\alpha)}(K_{1,n}^{{\rm v}(\alpha)})}\id,
\end{equation}
where $K_{1,n}^{{\rm v}(\alpha)} = \ln(\det{\bf F}^{{\rm v}(\alpha)}_n)$, and the viscous-volumetric-strain-dependent viscosities are calculated through \autoref{eq:eta_eq}.
\item[\bf Step 2.] Update the viscous distortion for mechanism $\alpha$ by using the forward Euler method to integrate \autoref{eq:evolution}:
\begin{equation}
\bfF^{{\rm v}(\alpha)}_{n+1} = (\id + \Delta t {\bf D}^{{\rm v}(\alpha)}_n){\bf F}^{{\rm v}(\alpha)}_n.
\end{equation}
\item[\bf Step 3.] Update the elastic distortion for mechanism $\alpha$:
\begin{equation}
\bfF^{{\rm e}(\alpha)}_{n+1}  =  \bfF_{n+1}\bfF^{{\rm v}(\alpha)-1}_{n+1}.
\end{equation}
\item[\bf Step 4.] Perform the polar decomposition and calculate the elastic logarithmic strain tensor for mechanism $\alpha$:
\begin{equation}
\bfF^{{\rm e}(\alpha)}_{n+1}=\bfV^{{\rm e}(\alpha)}_{n+1}\bfR^{{\rm e}(\alpha)}_{n+1}\quad\text{and}\quad \bfE^{{\rm e}(\alpha)}_{n+1}=\ln \bfV^{{\rm e}(\alpha)}_{n+1}.
\end{equation}
\item[\bf Step 5.] Calculate the non-equilibrium contribution to the Cauchy stress $\bfT_{n+1}^{(\alpha)}$ for mechanism $\alpha$ using \autoref{eq:cauchy_neq2}, \autoref{eq:coupled_fit}, \autoref{eq:L_1_neq}, and \autoref{eq:f_vr}.
\item[\bf Step 6.] Update the Mandel stress for non-equilibrium mechanism $\alpha$:
\begin{equation}
\bfM^{{\rm e}(\alpha)}_{n+1} = \left(\det\bfF^{{\rm e}(\alpha)}_{n+1}\right)\bfR_{n+1}^{{\rm e}(\alpha)}{}^\top\bfT_{n+1}^{(\alpha)} \bfR^{{\rm e}(\alpha)}_{n+1}.
\end{equation}
\end{enumerate}
The total Cauchy stress at time $t_{n+1}$ is then $\bfT_{n+1} = \bfT_{{\rm eq},n+1} + \sum_{\alpha=1}^{N}\bfT^{(\alpha)}_{n+1}$.

\bibliographystyle{elsarticle-harv} 
\bibliography{refs}

@article{flaschel2021unsupervised,
  title={Unsupervised discovery of interpretable hyperelastic constitutive laws},
  author={Flaschel, Moritz and Kumar, Siddhant and {De Lorenzis}, Laura},
  journal={Computer Methods in Applied Mechanics and Engineering},
  volume={381},
  pages={113852},
  year={2021},
  doi={10.1016/j.cma.2021.113852},
}

@phdthesis{landauer2019dissertation,
author={Landauer, Alexander},
title={Elastic Foam Characterization for Rate-Dependent Nonlinear Material Model Development},
school={Brown University},
year={2019},
}

@article{landauer2019quasistatic,
  title={Experimental characterization and hyperelastic constitutive modeling of open-cell elastomeric foams},
  author={Landauer, Alexander K and Li, Xiuqi and Franck, Christian and Henann, David L},
  journal={Journal of the Mechanics and Physics of Solids},
  volume={133},
  pages={103701},
  year={2019},
  doi = {10.1016/j.jmps.2019.103701},
  publisher={Elsevier}
}

@article{esposito2020modeling,
  title={Modeling of a Visco-Hyperelastic Polymeric Foam with a Continuous to Discrete Relaxation Spectrum Approach},
  author={Esposito, Marco and Sorrentino, Luigi and Krej{\v{c}}{\'\i}, Pavel and Davino, Daniele},
  journal={Journal of the Mechanics and Physics of Solids},
  pages={104030},
  volume = {142},
  year={2020},
  doi = {10.1016/j.jmps.2020.104030},
}

@article{koumlis2019strain,
  title={Strain Rate Dependent Compressive Response of Open Cell Polyurethane Foam},
  author={Koumlis, S and Lamberson, L},
  journal={Experimental Mechanics},
  volume={59},
  number={7},
  pages={1087--1103},
  year={2019},
  publisher={Springer},
  doi = {10.1007/s11340-019-00521-3},
}

@book{gibson97,
author = {Gibson, L. J. and Ashby, M. F.},
title = {Cellular solids: structures and properties, 2nd edn.},
publisher = {Cambridge University Press},
year = {1997}
}

@misc{abaqus,
author = {Abaqus/Explicit},
title = {Reference manuals},
year = {2026},
howpublished = {Dassault Syst\`{e}mes Simulia Corp.},
}

@article{briody2012,
author={Briody, C. and Duignan, B. and Jerrams, S. and Tiernan, J.},
journal ={Computational Material Science},
title = {The implementation of a visco-hyperelastic numerical material model for simulating the behaviour of polymer foam materials},
volume={64},
pages={47--51},
year={2012},
doi={10.1016/j.commatsci.2012.04.012}
}

@article{chen2012,
author = {W. Chen and {Z.-X.} Lu and B. Pan and {J.-H.} Guo and {Y.-J.} Li},
title = {Nonlinear behavior of bumper foams under uniaxial compressive cyclic loading},
journal = {Materials and Design},
volume = {35},
pages = {491--497},
year = {2012},
doi={10.1016/j.matdes.2011.09.042}
}

@article{anani2011,
author = {Y. Anani and Y. Alizadeh},
title = {Visco-hyperelastic constitutive law for modeling of foam's behavior},
journal = {Materials and Design},
volume = {32},
year = {2011},
pages = {2940--2948},
doi={10.1016/j.matdes.2010.11.010}
}

@article{yang2004,
author = {L. M. Yang and V. P. W. Shim},
title = {A visco-hyperelastic constitutive description of elastomeric foam},
journal = {International Journal of Impact Engineering},
volume = {30},
pages = {1099--1110},
year = {2004},
doi={10.1016/j.ijimpeng.2004.03.011}
}

@article{tang2017,
author = {M. Tang and G. Huang and H. Zhang and Y. Liu and H. Chang and H. Song and D. Xu and Z. Wang},
title = {Dependences of rheological and compression mechanical properties on cellular structures for impact-protective materials},
journal = {ACS Omega},
volume = {2},
year = {2017},
pages = {2214--2223},
doi={10.1021/acsomega.7b00242}
}

@article{ju2015,
author = {M. L. Ju and H. Jmal and R. Dupuis and E. Aubry},
title = {Visco-hyperelastic constitutive model for modeling the quasi-static behavior of polyurethane foam in large deformation},
journal = {Polymer Engineering and Science},
pages = {1795--1804},
year = {2015},
volume = {55},
doi={10.1002/pen.24018}
}

@article{ling2018,
title = {Deformation response of $\text{EPS}$ foam under combined compression-shear loading. Part I: Experimental design and quasi-static tests},
journal = {International Journal of Mechanical Sciences},
volume = {144},
pages = {480 -- 489},
year = {2018},
author = {Ling, C. and Ivens, J. and Cardiff, P. and Gilchrist, M. D.},
doi={10.1016/j.ijmecsci.2018.06.014}
}

@article{deshpande2001,
author = {V. S. Deshpande and N. A. Fleck},
title = {Multi-axial yield behaviour of polymer foams},
journal = {Acta Materialia},
volume = {49},
year = {2001},
pages = {1859--1866},
doi={10.1016/S1359-6454(01)00058-1}
}

@article{criscione2000,
  title={An invariant basis for natural strain which yields orthogonal stress response terms in isotropic hyperelasticity},
  author={Criscione, J. C. and Humphrey, J. D. and Douglas, A. S. and Hunter, W. C.},
  journal={Journal of the Mechanics and Physics of Solids},
  volume={48},
  pages={2445--2465},
  year={2000},
doi={10.1016/S0022-5096(00)00023-5}
}

@article{wang2002,
	title = {Full-field measurements of heterogeneous deformation patterns on polymeric foams using digital image correlation},
	volume = {39},
	doi = {10.1016/S0020-7683(02)00176-2},
	journal = {International Journal of Solids and Structures},
	author = {Y. Wang and A. M. {Cuiti\~no}},
	year = {2002},
	pages = {3777--3796}
}

@article{BarKochba2015,
  title={A fast iterative digital volume correlation algorithm for large deformations},
  author={{Bar-Kochba}, E. and Toyjanova, J. and Andrews, E. and Kim, {K.-S} and Franck, C.},
  journal={Experimental Mechanics},
  volume={55},
  pages={261--274},
  year={2015},
  doi={10.1007/s11340-014-9874-2}
}

@article{LI2022,
title = {Large-deformation constitutive modeling of viscoelastic foams: Application to a closed-cell foam material},
journal = {Journal of the Mechanics and Physics of Solids},
volume = {161},
pages = {104807},
year = {2022},
issn = {0022-5096},
doi = {https://doi.org/10.1016/j.jmps.2022.104807},
author = {Xiuqi Li and Jialiang Tao and Alexander K. Landauer and Christian Franck and David L. Henann},
}

@article{bergstrom1998constitutive,
  title={Constitutive modeling of the large strain time-dependent behavior of elastomers},
  author={Bergstr{\"o}m, Jorgen S and Boyce, M. C.},
  journal={Journal of the Mechanics and Physics of Solids},
  volume={46},
  number={5},
  pages={931--954},
  year={1998},
  doi={10.1016/S0022-5096(97)00075-6},
}

@article{mane2017,
	title = {Mechanical {Property} {Evaluation} of {Polyurethane} {Foam} under {Quasi}-static and {Dynamic} {Strain} {Rates}- {An} {Experimental} {Study}},
	volume = {173},
	issn = {1877-7058},
	doi = {10.1016/j.proeng.2016.12.160},
	journal = {Procedia Engineer},
	author = {Mane, J.V. and Chandra, S and Sharma, S and Ali, H and Chavan, V.M. and Manjunath, B.S. and Patel, R.J.},
	year = {2017},
	pages = {726--731}
}

@article{areias2017,
title = "Analysis of experimentally assessed EVA foams with mixed solid-shell elements capable of very large strains",
journal = "Finite Elements in Analysis and Design",
volume = "128",
pages = "19 - 31",
year = "2017",
issn = "0168-874X",
doi = "https://doi.org/10.1016/j.finel.2017.01.003",
author = {P. Areias and A. Rodrigues and T. Rabczuk and J. Garcao and A. Carvalho}
}

@article{sun2018,
	title = {Dynamic compressive behaviour of cellular materials: a review of phenomenon, mechanism and modelling},
	author = {Sun, Y. and Li, Q.M.},
	year = {2017},
	volume = {112},
    pages = {74--115},
	journal = {International Journal of Impact Engineering},
	doi={10.1016/j.ijimpeng.2017.10.006}
}

@article{kossa2015,
  title={A new biaxial compression fixture for polymeric foams},
  author={Kossa, Attila},
  journal={Polymer Testing},
  volume={45},
  pages={47--51},
  year={2015},
  publisher={Elsevier},
  doi={10.1016/j.polymertesting.2014.08.003}
}

@article{landauer2018qdic,
author={Landauer, A. K. and Patel, M. and Henann, D. L. and Franck, C.},
title={A q-factor-based digital image correlation algorithm {($\text{qDIC}$)} for resolving finite deformations with degenerate speckle patterns},
journal={Experimental Mechanics},
year={2018},
volume={58},
pages={815--830},
doi={10.1007/s11340-018-0377-4}
}

@Article{liu2005,
author="Liu, Qunli
and Subhash, Ghatu
and Gao, Xin-Lin",
title="A Parametric Study on Crushability of Open-Cell Structural Polymeric Foams",
journal="Journal of Porous Materials",
year="2005",
volume="12",
number="3",
pages="233--248",
issn="1573-4854",
doi="10.1007/s10934-005-1652-1"
}

@article{okolieocha2015,
  title={Microcellular to nanocellular polymer foams: {Progress} (2004--2015) and future directions--{A review}},
  author={Okolieocha, Chimezie and Raps, Daniel and Subramaniam, Kalaivani and Altst{\"a}dt, Volker},
  journal={European Polymer Journal},
  volume={73},
  pages={500--519},
  year={2015},
  doi={10.1016/j.eurpolymj.2015.11.001}
}

@article{azizi2016,
  title={Identification of nonlinear viscoelastic models of flexible polyurethane foam from uniaxial compression data},
  author={Azizi, Yousof and Davies, Patricia and Bajaj, Anil K},
  journal={Journal of Engineering Materials and Technology},
  volume={138},
  number={2},
  pages={021008},
  year={2016},
  doi={10.1115/1.4032169}
}

@article{xi2017,
	title = {Meso-scale mechanism of compaction shock propagation in cellular materials},
	volume = {109},
	issn = {0734-743X},
	doi = {10.1016/j.ijimpeng.2017.07.005},
	journal = {International Journal of Impact Engineering},
	author = {Xi, C. Q. and Li, Q. M.},
	year = {2017},
    pages = {321--334}
}

@article{piero2012,
	title = {The influence of viscosity on the response of open-cell polymeric foams in uniaxial compression: experiments and theoretical model},
	volume = {24},
	issn = {0935-1175},
	doi = {10.1007/s00161-011-0230-9},
	number = {3},
	journal = {Continuum Mech Therm},
	author = {{Del Piero}, Gianpietro and Pampolini, Giampiero},
	year = {2012},
	pages = {181--199}
}

@article{viot2009,
	title = {Hydrostatic compression on polypropylene foam},
	volume = {36},
	issn = {0734-743X},
	doi = {10.1016/j.ijimpeng.2008.11.010},
    number = {7},
	journal = {International Journal of Impact Engineering},
	author = {Viot, Philippe},
	year = {2009},
	pages = {975--989}
}

@article{roux2008,
	title = {Three-dimensional image correlation from {X}-ray computed tomography of solid foam},
	volume = {39},
	doi = {10.1016/j.compositesa.2007.11.011},
	number = {8},
	journal = {Composites Part A: Applied Science and Manufacturing},
	author = {Roux, Stephane and Hild, Francois and Viot, Philippe and Bernard, Dominique},
	year = {2008},
	pages = {1253--1265}
}

@article{mills2009,
	title = {Finite element micromechanics model of impact compression of closed-cell polymer foams},
	volume = {46},
	issn = {0020-7683},
	doi = {10.1016/j.ijsolstr.2008.09.012},
    number = {3},
	journal = {International Journal of Solids and Structures},
	author = {Mills, N. J. and Stampfli, R. and Marone, F. and Bruhwiler, P. A.},
	year = {2009},
	pages = {677--697}
}

@article{sun2016critshock,
	title = {Determination of the constitutive relation and critical condition for the shock compression of cellular solids},
	volume = {99},
	issn = {0167-6636},
	doi = {10.1016/j.mechmat.2016.04.004},
	journal = {Mechanics of Materials},
	author = {Sun, Yongle and Li, Q. M. and McDonald, S. A. and Withers, P. J.},
	month = aug,
	year = {2016},
	pages = {26--36}
}

@article{markert2008,
	title = {A {Biphasic} {Continuum} {Approach} for {Viscoelastic} {High}-{Porosity} {Foams}: {Comprehensive} {Theory}, {Numerics}, and {Application}},
	volume = {15},
	issn = {1134-3060},
	doi = {10.1007/s11831-008-9023-0},
	number = {4},
	journal = {Arch Comput Method E},
	author = {Markert, Bernd},
	year = {2008},
	pages = {371--446}
}

@article{jeong2016,
	title = {Constitutive modeling of polymeric foams having a four-parameter modulus function with strain rate sensitivity},
	volume = {30},
	issn = {1976-3824},
	doi = {10.1007/s12206-016-0124-7},
    journal = {Journal of Mechanical Science and Technology},
	author = {Jeong, Kwang Young},
	year = {2016},
    pages = {683--688},
}

@article{rusch1969,
	title = {Load-compression behavior of flexible foams},
	volume = {13},
	doi = {10.1002/app.1969.070131106},
	number = {11},
	journal = {Journal of Applied Polymer Science},
	author = {Rusch, K. C.},
	year = {1969},
	pages = {2297--2311}
}

@article{sahraoui2000,
	title = {Mechanical testing of polymeric foams at low frequency},
	volume = {20},
	issn = {0142-9418},
	doi = {10.1016/S0142-9418(00)00006-4},
	number = {1},
	journal = {Polymer Testing},
	author = {Sahraoui, S and Mariez, E and Etchessahar, M},
	year = {2000},
	pages = {93--96}
}

@article{yi2006,
	title = {Large deformation rate-dependent stress-strain behavior of polyurea and polyurethanes},
	volume = {47},
	issn = {0032-3861},
	doi = {10.1016/j.polymer.2005.10.107},
	number = {1},
	journal = {Polymer},
	author = {Yi, J. and Boyce, M. C. and Lee, G. F. and Balizer, E.},
	year = {2006},
	pages = {319--329}
}

@article{tao2024experimental,
  title={Experimental approach for characterizing the nonlinear, time and temperature-dependent constitutive response of open-cell polyurethane foams},
  author={Tao, Jialiang and Sun, Xiangyu and Franck, Christian},
  journal={Strain},
  volume={60},
  pages={e12478},
  year={2024},
  doi={10.1111/str.12478},
}

@article{kossa2016,
	title = {Visco-hyperelastic {Characterization} of {Polymeric} {Foam} {Materials}},
	volume = {3},
	issn = {2214-7853},
	doi = {10.1016/j.matpr.2016.03.037},
	number = {4},
	journal = {Mater Today Proc},
	author = {Kossa, Attila and Berezvai, Szabolcs},
	year = {2016},
	pages = {1003--1008}
}

@article{ma2009,
	title = {Modeling loading rate effect on crushing stress of metallic cellular materials},
	volume = {36},
	issn = {0734-743X},
	doi = {10.1016/j.ijimpeng.2008.11.013},
	number = {6},
	journal = {International Journal of Impact Engineering},
	author = {Ma, G. W. and Ye, Z. Q. and Shao, Z. S.},
	year = {2009},
	pages = {775--782}
}

@article{farid2004,
	title = {Differentiation of discrete multidimensional signals},
	volume = {13},
	issn = {1057-7149},
	doi = {10.1109/TIP.2004.823819},
	number = {4},
	journal = {IEEE Transactions on Image Processing},
	author = {Farid, H. and Simoncelli, E. P.},
	year = {2004},
	pages = {496--508}
}

@article{calladine1984,
	title = {Strain-rate and inertia effects in the collapse of two types of energy-absorbing structure},
	volume = {26},
	issn = {0020-7403},
	doi = {10.1016/0020-7403(84)90021-3},
	number = {11},
	journal = {International Journal of Mechanical Sciences},
	author = {Calladine, C. R. and English, R. W.},
	year = {1984},
	pages = {689--701}
}

@article{shen2010,
	title = {Compressive behaviour of closed-cell aluminium foams at high strain rates},
	volume = {41},
	issn = {1359-8368},
	doi = {10.1016/j.compositesb.2010.07.005},
	journal = {Composites Part B: Engineering},
	author = {Shen, Jianhu and Lu, Guoxing and Ruan, Dong},
	year = {2010},
	pages = {678--685}
}

@book{mills2007polymer,
  title={Polymer foams handbook: engineering and biomechanics applications and design guide},
  author={Mills, Nigel},
  year={2007},
  publisher={Elsevier}
}

@article{landauer2023dataframework,
  title={A materials data framework and dataset for elastomeric foam impact mitigating materials},
  author={Landauer, Alexander K and Kafka, Orion L and Moser, Newell H and Foster, Ian and Blaiszik, Ben and Forster, Aaron M},
  journal={Scientific Data},
  volume={10},
  number={1},
  pages={356},
  year={2023},
  doi = {10.1038/s41597-023-02092-4},
  publisher={Nature Publishing Group UK London}
}

@article{koohbor_multiscale_2021,
	title = {A multiscale experimental approach to characterize micro-to-macro transition length scale in polymer foams},
	volume = {161},
	issn = {0167-6636},
	doi = {10.1016/j.mechmat.2021.104006},
	journal = {Mechanics of Materials},
	author = {Koohbor, Behrad and Pagliocca, Nicholas and Youssef, George},
	year = {2021},
	pages = {104006}
}

@article{mills_modelling_2002,
	title = {Modelling of Air Flow in Impacted Polyurethane Foam},
	volume = {21},
	issn = {0262-4893},
	doi = {10.1177/026248930202100502},
	number = {5},
	journal = {Cellular Polymers},
	author = {Mills, N.J. and Lyn, G.},
	year = {2002},
	pages = {343--368}
}

@article{rouwane_architecture-driven_2022,
	title = {Architecture-{Driven} {Digital} {Image} {Correlation} {Technique} ({ADDICT}) for the measurement of sub-cellular kinematic fields in speckle-free cellular materials},
	volume = {234-235},
	issn = {0020-7683},
	doi = {10.1016/j.ijsolstr.2021.111223},
	journal = {International Journal of Solids and Structures},
	author = {Rouwane, Ali and Bouclier, Robin and Passieux, Jean-Charles and Périé, Jean-Noël},
	year = {2022},
	pages = {111223},
}

@article{JOSHI2026118719,
title = {Instabilities and phase transitions in architected metamaterials: a gradient-enhanced continuum approach},
journal = {Computer Methods in Applied Mechanics and Engineering},
volume = {452},
pages = {118719},
year = {2026},
doi = {https://doi.org/10.1016/j.cma.2025.118719},
author = {Sarvesh Joshi and S. Mohammad Mousavi and Craig M. Hamel and Stavros Gaitanaros and Prashant K. Purohit and Ryan Alberdi and Nikolaos Bouklas},
}

@article{bhagavathula_density_2022,
	title = {Density, strain rate and strain effects on mechanical property evolution in polymeric foams},
	volume = {161},
	issn = {0734-743X},
	doi = {10.1016/j.ijimpeng.2021.104100},
	journal = {International Journal of Impact Engineering},
	author = {Bhagavathula, Kapil Bharadwaj and Meredith, Christopher S and Ouellet, Simon and Romanyk, Dan L and Hogan, James David},
	year = {2022},
	pages = {104100}
}

@article{landauer_stress_2020,
	title = {Stress and Strain Heuristics for a Layered Elastomeric Foam at Medium Impact Rates},
	url = {https://www.nist.gov/publications/stress-and-strain-heuristics-layered-elastomeric-foam-medium-impact-rates},
	journal = {NIST},
	author = {Landauer, Alexander and Blitterswyk, Jared C. Van and Riley, Michael A. and Forster, Aaron M.},
	year = {2020}
}

@article{bailly_strain_2020,
	title = {Strain Rate Dependent behavior of Vinyl Nitrile Helmet Foam in Compression and Combined Compression and Shear},
	volume = {10},
	issn = {2076-3417},
	doi = {10.3390/app10228286},
	number = {22},
	urldate = {2023-06-15},
	journal = {Applied Sciences},
	author = {Bailly, Nicolas and Petit, Yvan and Desrosier, Jean-Michel and Laperriere, Olivier and Langlois, Simon and Wagnac, Eric},
	month = jan,
	year = {2020},
	pages = {8286}
}

@article{morrison2024temperature,
  title={Temperature dependent dynamic response of open-cell polyurethane foams},
  author={Morrison, D and Morton, J and Foster, M and Lamberson, L},
  journal={Experimental Mechanics},
  volume={64},
  pages={929--943},
  year={2024},
  doi = {10.1007/s11340-024-01054-0},
}

@misc{localization_code,
    title = {Code for qDIC-based Strain Localization Analysis with Wavelets (Q-SLAW)},
    author = {Landauer, Alexander K},
    year = {2024},
    doi = {10.18434/mds2-3141},
    publisher = {National Institute of Standards and Technology},
    urldate = {2024-01-10}
}

@article{HUAJIAN19922471,
title = {Elastic contact versus indentation modeling of multi-layered materials},
journal = {International Journal of Solids and Structures},
volume = {29},
number = {20},
pages = {2471-2492},
year = {1992},
issn = {0020-7683},
doi = {10.1016/0020-7683(92)90004-D},
author = {Gao Huajian and Chiu Cheng-Hsin and Lee Jin},
}

@article{yan2026equilibrium,
  title={{An Equilibrium Constrained Genetic Algorithm (ECGA) calibration framework for isotropic hyperelastic constitutive models with application to an elastomeric foam material}},
  author={Yan, Zicheng and Tao, Jialiang and Franck, Christian and Henann, David L},
  journal={Computer Methods in Applied Mechanics and Engineering},
  volume={457},
  pages={118978},
  year={2026},
  doi = {10.1016/j.cma.2026.118978},
  publisher={Elsevier}
}

@article{tacc2023data,
  title={Data-driven anisotropic finite viscoelasticity using neural ordinary differential equations},
  author={Ta{\c{c}}, Vahidullah and Rausch, Manuel K and Costabal, Francisco Sahli and Tepole, Adrian Buganza},
  journal={Computer Methods in Applied Mechanics and Engineering},
  volume={411},
  pages={116046},
  year={2023},
  publisher={Elsevier},
  doi={10.1016/j.cma.2023.116046}
}

@article{abdolazizi2024viscoelastic,
  title={Viscoelastic constitutive artificial neural networks (vCANNs)--A framework for data-driven anisotropic nonlinear finite viscoelasticity},
  author={Abdolazizi, Kian P and Linka, Kevin and Cyron, Christian J},
  journal={Journal of computational physics},
  volume={499},
  pages={112704},
  year={2024},
  doi={10.1016/j.jcp.2023.112704},
  publisher={Elsevier}
}

@article{kalina2026physics,
  title={A physics-augmented neural network framework for finite strain incompressible viscoelasticity},
  author={Kalina, Karl A and Brummund, J{\"o}rg and K{\"a}stner, Markus},
  journal={Computer Methods in Applied Mechanics and Engineering},
  volume={455},
  pages={118892},
  year={2026},
  doi={10.1016/j.cma.2026.118892},
  publisher={Elsevier}
}

@article{liang2008neural,
  title={Neural network based constitutive model for elastomeric foams},
  author={Liang, Guanghui and Chandrashekhara, K},
  journal={Engineering Structures},
  volume={30},
  number={7},
  pages={2002--2011},
  year={2008},
  publisher={Elsevier},
  doi={10.1016/j.engstruct.2007.12.021}
}

@article{frankel2022machine,
  title={Machine learning constitutive models of elastomeric foams},
  author={Frankel, Ari and Hamel, Craig M and Bolintineanu, Dan and Long, Kevin and Kramer, Sharlotte},
  journal={Computer Methods in Applied Mechanics and Engineering},
  volume={391},
  pages={114492},
  year={2022},
  publisher={Elsevier},
  doi={10.1016/j.cma.2021.114492}
}

@article{williams_temperature_1955,
	title = {The {Temperature} {Dependence} of {Relaxation} {Mechanisms} in {Amorphous} {Polymers} and {Other} {Glass}-forming {Liquids}},
	volume = {77},
	issn = {0002-7863},
	doi = {10.1021/ja01619a008},
	number = {14},
	journal = {Journal of the American Chemical Society},
	publisher = {American Chemical Society},
	author = {Williams, Malcolm L. and Landel, Robert F. and Ferry, John D.},
	year = {1955},
	pages = {3701--3707}
}

\end{document}